%% file: Betelgeuse_seismo052518.tex
\documentclass[fleqn,usenatbib,useAMS]{mnras}
\usepackage{graphicx,xfrac,amsmath,longtable,color}
\usepackage{float}
\usepackage[caption = false]{subfig}
\usepackage{rotating}
\usepackage{pdflscape}
\include{definitions}

\usepackage[T1]{fontenc}
\usepackage{ae,aecompl}

\title{The Betelgeuse Project II: Asteroseismology}

\author[S.Nance et al.]{
S. Nance,$^{1}$ J. M. Sullivan,$^{1}$ M. Diaz,$^{1}$
J. Craig Wheeler,$^{1}$\thanks{Contact e-mail: wheel@astro.as.utexas.edu}  
 \\
$^{1}$Department of Astronomy, University of Texas at Austin, Austin, TX, USA\\
}

\begin{document}

\pubyear{2017}

\label{firstpage}
\pagerange{\pageref{firstpage}--\pageref{lastpage}}
\maketitle

\begin{abstract}

We explore the question of whether the interior state of massive red supergiant 
supernova progenitors can be effectively probed with asteroseismology. We 
have computed a suite of ten models with ZAMS masses from 15 to 25 \msun\ 
in intervals of 1 \msun\, including the effects of rotation, with the stellar evolutionary 
code {\it MESA}. We estimate characteristic frequencies and convective luminosities 
of convective zones at two illustrative stages, core helium burning and off-center 
convective carbon burning. We also estimate the power that might be delivered to 
the surface to modulate the luminous output considering various efficiencies and 
dissipation mechanisms. The inner convective regions should generate waves 
with characteristic periods of $\sim$ 20 days in core helium burning, $\sim$10 
days in helium shell burning, and 0.1 to 1 day in shell carbon burning. Acoustic waves 
may avoid both shock and diffusive dissipation relatively early in core helium burning 
throughout most of the structure. In shell carbon burning, years before explosion, the 
signal generated in the helium shell might in some circumstances be weak enough to 
avoid shock dissipation, but is subject to strong thermal dissipation in the hydrogen
envelope. Signals from a convective carbon-burning shell are very likely to be even 
more severely damped by within the envelope. In the most optimistic case, early in 
core helium burning, waves arriving close to the surface could represent luminosity 
fluctuations of a few millimagnitudes, but the conditions in the very outer reaches of 
the envelope suggest severe thermal damping there.

%248 words
\end{abstract}

\begin{keywords}
stars: individual (Betelgeuse) --- stars: evolution --- stars: AGB 
and post-AGB --- supernovae: general
\end{keywords}

%TOP

\section{Introduction}
\label{intro}

Betelguese ($\alpha$ Orionis) is a massive red supergiant (RSG) that is destined to 
explode as a Type IIP supernova (some speculate explosion as a blue supergiant)
and leave behind a neutron star. It is thus a touchstone for a broad range of issues 
of the evolution and explosion of massive stars. One of the outstanding issues is 
that we do not have tight constraints on the evolutionary state of Betelgeuse and 
hence when it might explode. Closely related issues are uncertainties about the 
internal rotational state and associated mixing. Understanding Betelgeuse in greater 
depth will yield benefits to the entire subject of massive star evolution.

As commanding as its presence is in the evening sky, its brightness limits certain 
aspects of the study of Betelgeuse. The distance is currently known to only 20\% 
($D = 197\pm45$ pc; Harper et al. 2008, 2017), and it is too bright to be observed by 
{\sl Gaia}. This leaves key properties such as radius and luminosity somewhat uncertain. 
On the other hand, its closeness allows other key measurements since its image can be 
resolved (Haubois et al. 2009). A particularly interesting potential constraint on 
Betelgeuse is the rotational velocity ($\sim 15$~\kms) measured with $HST$ 
\citet{Dupree87}. Betelgeuse shows a 420 day period  \citet{Dupree87} that is most 
likely a first overtone radial pulsation mode that has not been broadly exploited as a 
constraint on Betelgeuse. It also shows variance on timescales of 2000 days that is
associated with overturn of convective plumes. Many apparently single massive stars 
have undergone mergers. Could that be true of Betelgeuse? 

For a judicious choice of distance, Betelgeuse might be brought into agreement 
with observations of $L$, $R$, and $T_{eff}$ at either the minimum--luminosity 
base of the red supergiant branch (RSB) or at the tip of the RSB. Single--star models 
give a rotational velocity of about that observed only near the minimum luminosity 
with the velocity at the tip of the giant branch being far below the suggested observed 
value. The former seems very improbable, and this phase disagrees with estimates
of log $g$ and with the possible observed rate of contraction of the radius of 
Betelgeuse. The latter comports with standard assumptions, but cannot easily 
account for the claimed rotational velocity. In the first paper of {\sl The Betelgeuse 
Project} \citep{Wheeler17}, we showed that single-star models have difficulty 
accounting for the rapid equatorial rotation and suggested that Betelgeuse might 
have merged with a companion of about 1 \msun\ to provide the requisite angular 
momentum to the envelope. Recent observations of Betelgeuse with {\sl ALMA}
are consistent with a rapid equatorial velocity \citep{Kervella18}. This reinforces the
notion that the results of \citet{Dupree87} are due to systematic rotational motion
rather than the motion of convective plumes and hence the need to understand this 
remarkably high equatorial velocity. Nevertheless, the convective plumes are real,
as revealed also in the RSG Antares \citep{Montarges17}, and Betelgeuse shows a
variety of complex motion on its surface that calls for deeper understanding before 
the rotational velocity can be solidly confirmed.

We need another means to determine the inner evolutionary state of Betelgeuse and, 
by extension, any massive RSG. Specifically, to resolve the uncertainties of the mass 
and evolutionary state of Betelgeuse, we need to peer inside. The most logical tool is 
asteroseismology \citep{Aerts15}. Stellar structure and evolution still is mostly 
pursued with spherical models not far removed from pioneering work by Hoyle, 
Schwarzschild and others. We should be prepared to find that the innards of real 
3D RSG are interestingly different, as we have found for the Sun. 

This work is an extension of \citet{Wheeler17} and hence presented in the context 
of models for Betelgeuse, but the principles should apply to RSG more broadly. 
Section \ref{comp} presents our model evolutionary computations, \S \ref{seismo} 
gives our results for characteristic frequencies generated near the base and the 
tip of the RSB, \S \ref{detect} addresses the possibility 
of detecting evidence of these waves at the surface, and \S \ref{discuss} 
presents a discussion of our results.

\section{Computations}
\label{comp}

\begin{figure}
\center
\subfloat[]{\includegraphics[width = 3.0in, angle=0]{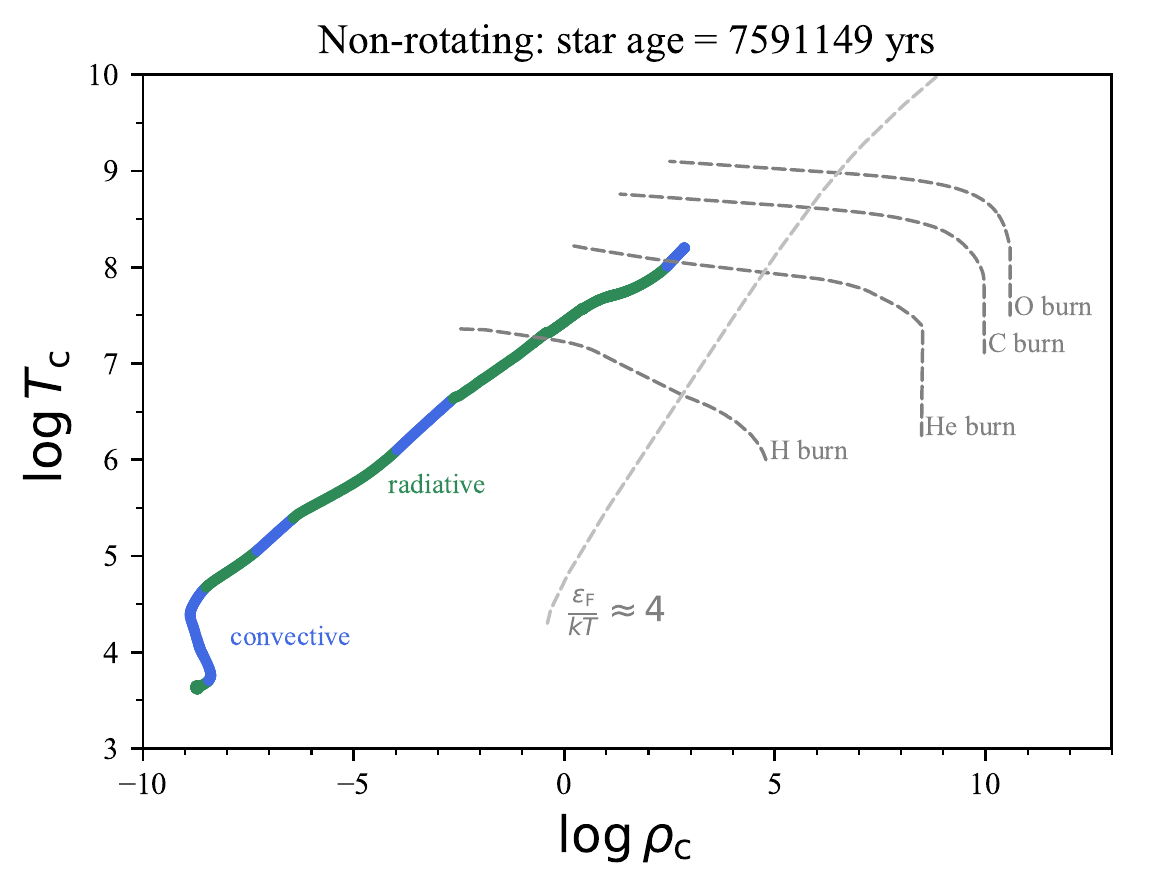}}
\center
\subfloat[]{\includegraphics[width = 3.0in, angle=0]{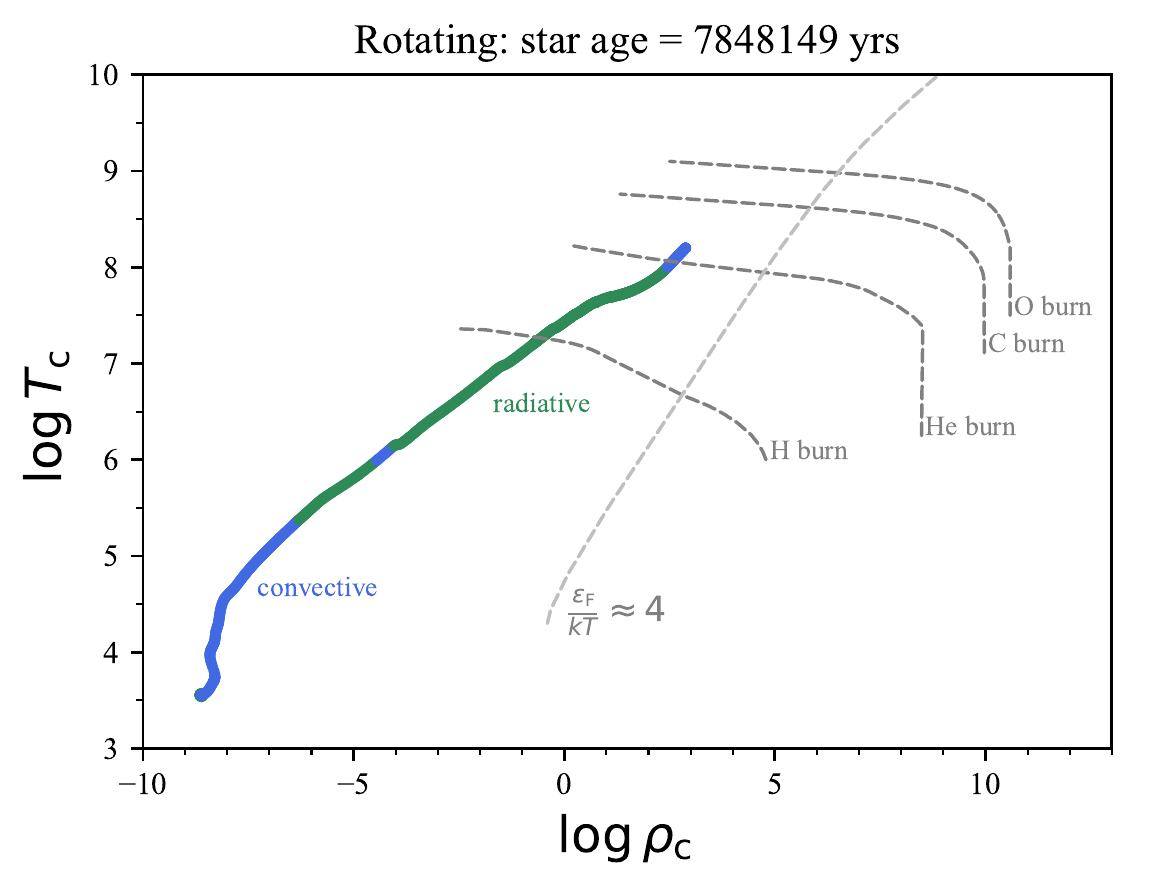}}
\caption{Structure of the 20~\m\ model in the temperature/density plane in core 
helium burning near the luminosity minimum for the non-rotating model (panel a) 
and for the rotating model (panel b). The blue regions are convective, the 
green regions, radiative. The dashed line corresponding to $\epsilon_F/(kT)\approx4$
represents the non-degenerate/degenerate boundary. Note in both models the convective 
region around $10^6$ K that is separated from the outer convective envelope by a 
radiative region.}
\label{trhohelium}
\end{figure}

\begin{figure}
\center
\subfloat[]{\includegraphics[width = 3.0in, angle=0]{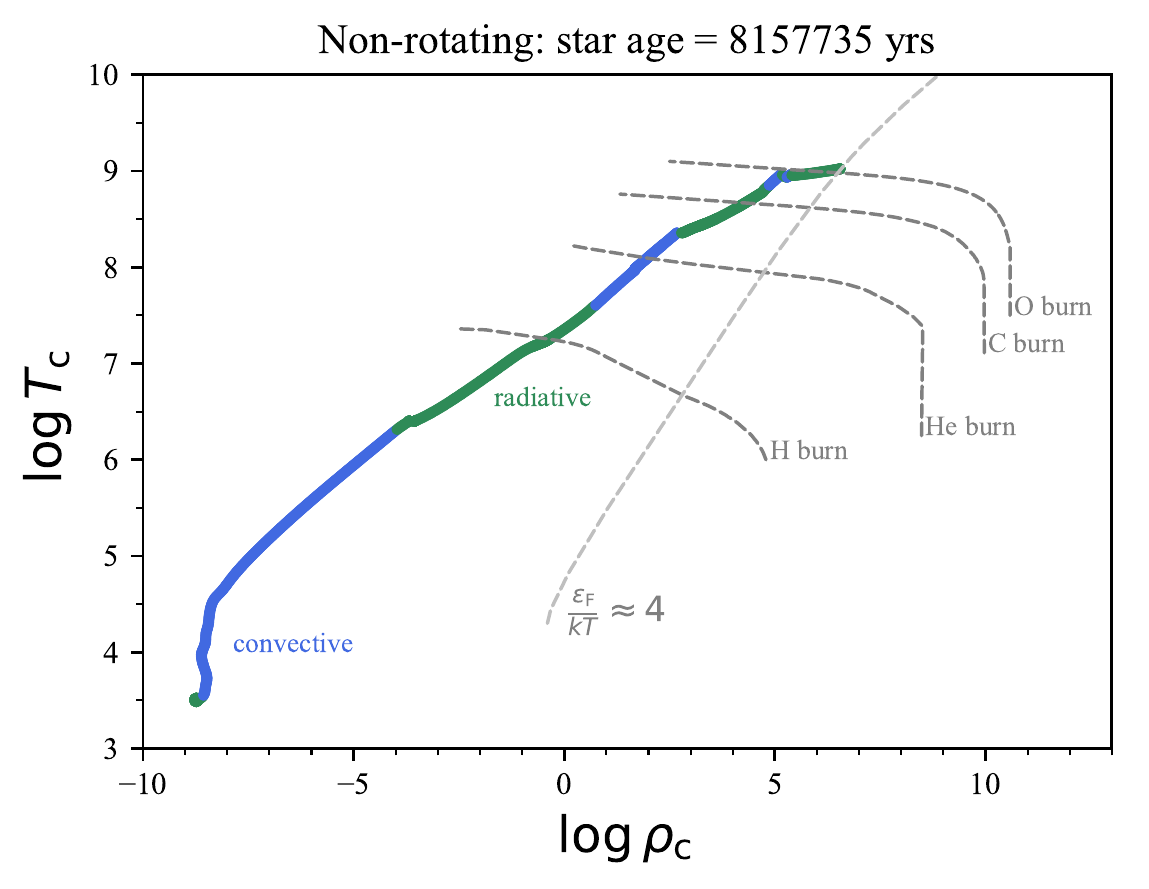}}
\center
\subfloat[]{\includegraphics[width = 3.0in, angle=0]{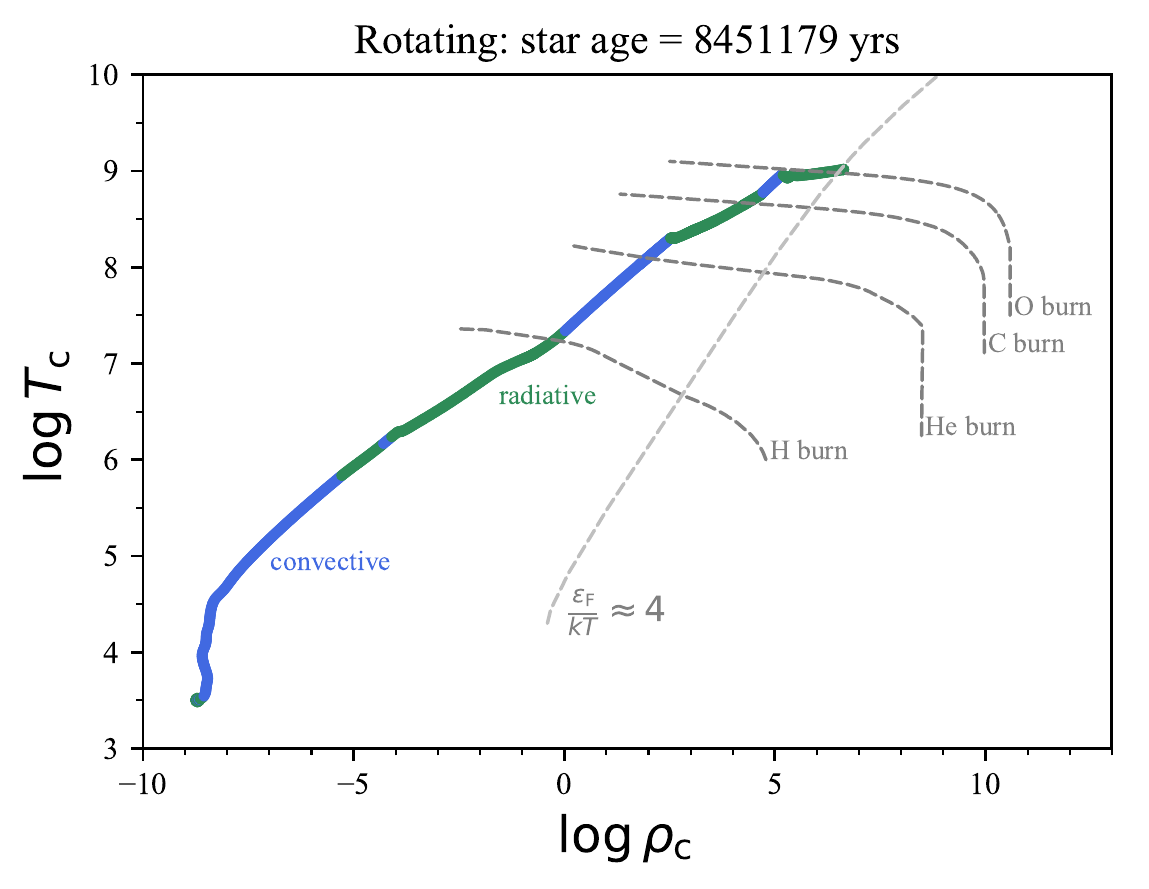}}
\caption{Structure of the 20~\m\ model in the temperature/density plane in the phase of
off-center convective carbon burning for the non-rotating model (panel a) and for
the rotating model (panel b). The blue regions are convective, the green regions, radiative.}
\label{trhocarbon}
\end{figure}

We evolved a grid of models from the Zero Age Main Sequence (ZAMS) to near the 
onset of core collapse using the stellar evolution code Modules for Experiments in 
Stellar Astrophysics ({\it MESA}; \citealt{Paxton11, Paxton13}). 
We computed models of solar metallicity with ZAMS masses from 15 to 25 \msun\
at intervals of 1 \msun\ primarily using {\it MESA} version 6208. Most of the rotating
models employed version 7624. One suite of models were non--rotating and another 
suite with the same ZAMS masses began with an initial rotation of $\sim$ 200~\kms, 
corresponding to about 25\% of the Keplerian velocity on the ZAMS. Because our 
principal goal was to explore the promise of an asterosesimological probe, we 
chose only the default prescriptions in {\it MESA}, Schwarzschild convection 
and an overshoot parameter of $\alpha = 0.2$. For the rotating models, we again 
chose  {\it MESA} default values of mechanisms of angular momentum transport 
and mixing. We did not include magnetic effects \citep{WKC15}. 
We employed the ``Dutch" mass--loss prescriptions with $\eta = 0.8$. We used 
nuclear reaction network {\sl approx21}. The inlist we employed is available 
upon request from the authors. For each ZAMS mass, the models were computed 
to the onset of core collapse. We present here data during early core helium burning
at the base of the RSB and at the first episode of off-center carbon shell burning. 
The formal criterion we adopted in the latter case was that the central mass fraction
of carbon be less than $10^{-5}$. By that criterion, a couple of the models were
in a phase without convective carbon burning. We then chose models at slightly
different times such that they did have off-center convective carbon burning
for the sake of homogeneity of sampling the associated convective frequencies.
Given the richness of the interior structure of massive stars and its sensitive variation
with time, these choices are purely illustrative. Figure \ref{trhohelium} illustrates 
the structure of the helium-burning stage for the non-rotating and rotating models 
of 20 \msun\ and Figure \ref{trhocarbon} the non-rotating and rotating models
in off-center carbon shell burning. Rotation as we have invoked it has relatively 
little effect on the inner structure. In the later stages there is relatively little 
difference in the outer, observable, characteristics as the models evolve 
\citep{Wheeler17}. There are significant differences in the inner convective motions 
that change rapidly with time that might give asteroseismological clues to the 
evolutionary state. In the following discussion, we refer to specific models by 
their ZAMS mass.

Figure \ref{trhohelium} and panel (b) of Figure \ref{trhocarbon} are 
characteristic of many of the models by showing a convective region around 
$10^6$ K that is separated from the outer convective envelope by a radiative region.
This convective region could generate gravity waves and acoustic flux, but the
frequencies tend to be low, so such signals are unlikely to propagate to the surface
(\S \ref{seismo}). The ``radiative" region around $10^5$ K in panel (a) of Figure \ref{trhohelium} 
is ephemeral with low convective velocity. In our work, we have basically classified 
this region as convective; this has little impact on our results.

\section{Characteristic Frequencies}
\label{seismo}

We estimated characteristic frequencies that might send acoustic signals to 
the surface. While it is conceivable that Betelgeuse is subject to opacity or 
nuclear--driven p--mode and g--mode pulsations and that this is the origin
of the 420 day pulsation (\S \ref{discuss}), for this work we made estimates 
guided by the notion that the inner regions of a RSG will be characterized by 
turbulent noise associated with strong convection in the late stages of evolution 
as explored by \citet{Arnett11}, \citet{SmithArnett14}, and \citet{CCAT15}. 
Several works have employed these notions to explore the possibility that
strong acoustic flux generated late in the evolution can lead to envelope expansion
or even mass loss \citep{QS12,SQ14,Fuller17}. Here we address earlier, gentler
phases in an attempt to seek clues to the the evolutionary state of RSG prior
to the final collapse. 

We followed the techniques and methods outlined by 
\citet{SQ14} and \citet{Fuller17} to determine convective and cutoff frequencies
and dissipative processes. $MESA$ computes the distribution of convective 
velocities in convective regions. The distribution is not symmetric about 
the midpoint of typical convective zones, but nearly so. We thus estimated 
a characteristic angular frequency for the overturn of a convective eddy as
\begin{equation}
\omega_{conv} \approx \frac{v_{conv}}{H_p}
\label{char_freq}
\end{equation}
where $v_{conv}$ is the convective velocity in the middle of a convective core
or shell, determined to be the half--way point in radius between boundaries of
a shell or half the radius of a central convective core, and ${H_p}$ is the
pressure scale height at the radial mid--point of the convective core or shell. 
A characteristic timescale associated with this frequency is thus
\begin{equation}
T_{conv} =  \frac{2\pi}{\omega_{conv}}  \approx \frac{2\pi H_p}{v_{conv}}.
\label{time}
\end{equation}
We note that \citet{Fuller17} defines the convective frequency to be 
$2\pi v_{conv}/2\alpha H_p$ with $\alpha = 2$ and hence larger than adopted 
here and by \citet{SQ14} by a factor of $\pi/2$. 

\begin{figure}
\center
\subfloat[]{\includegraphics[width = 3.0in, angle=0]{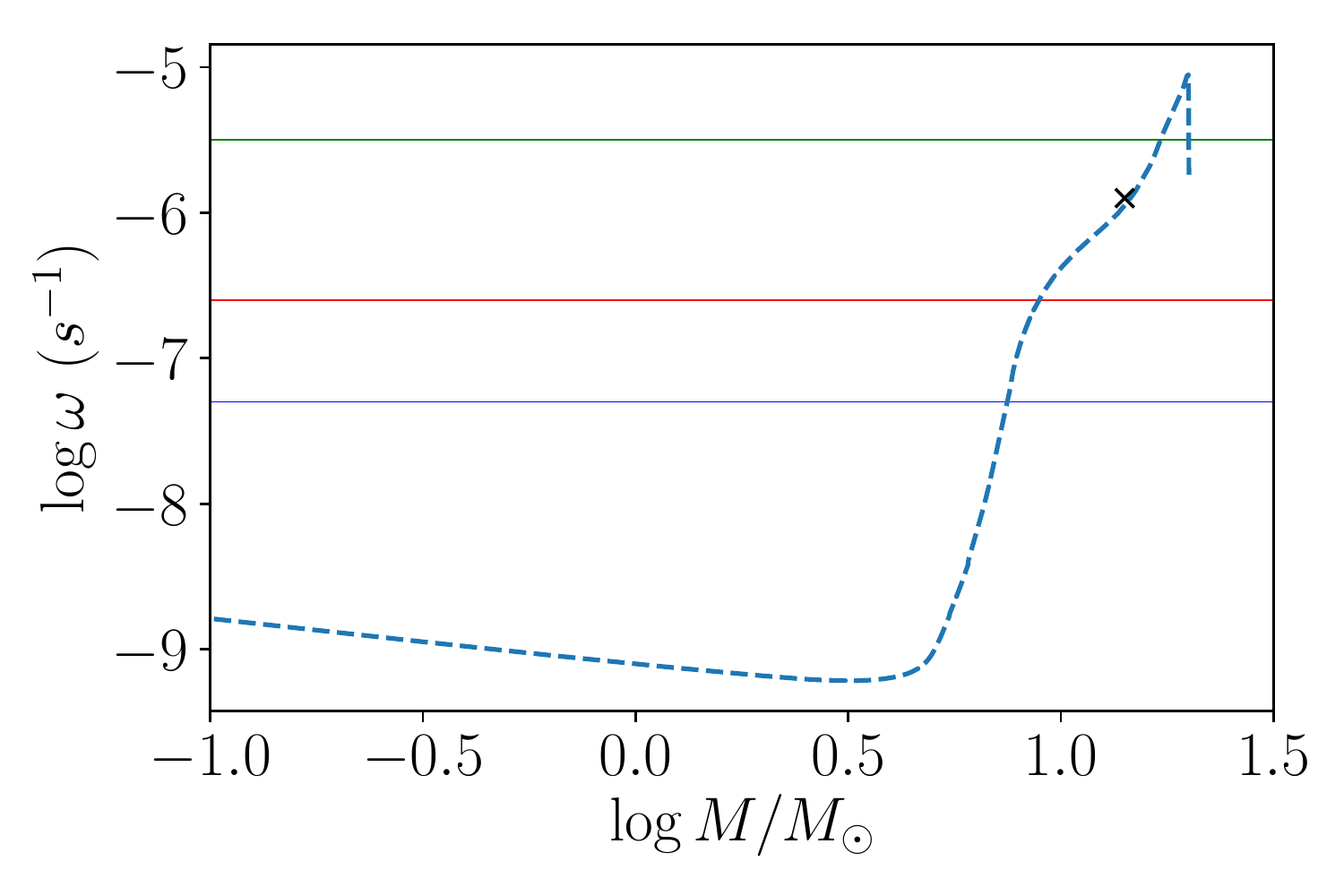}}
\center
\subfloat[]{\includegraphics[width = 3.0in, angle=0]{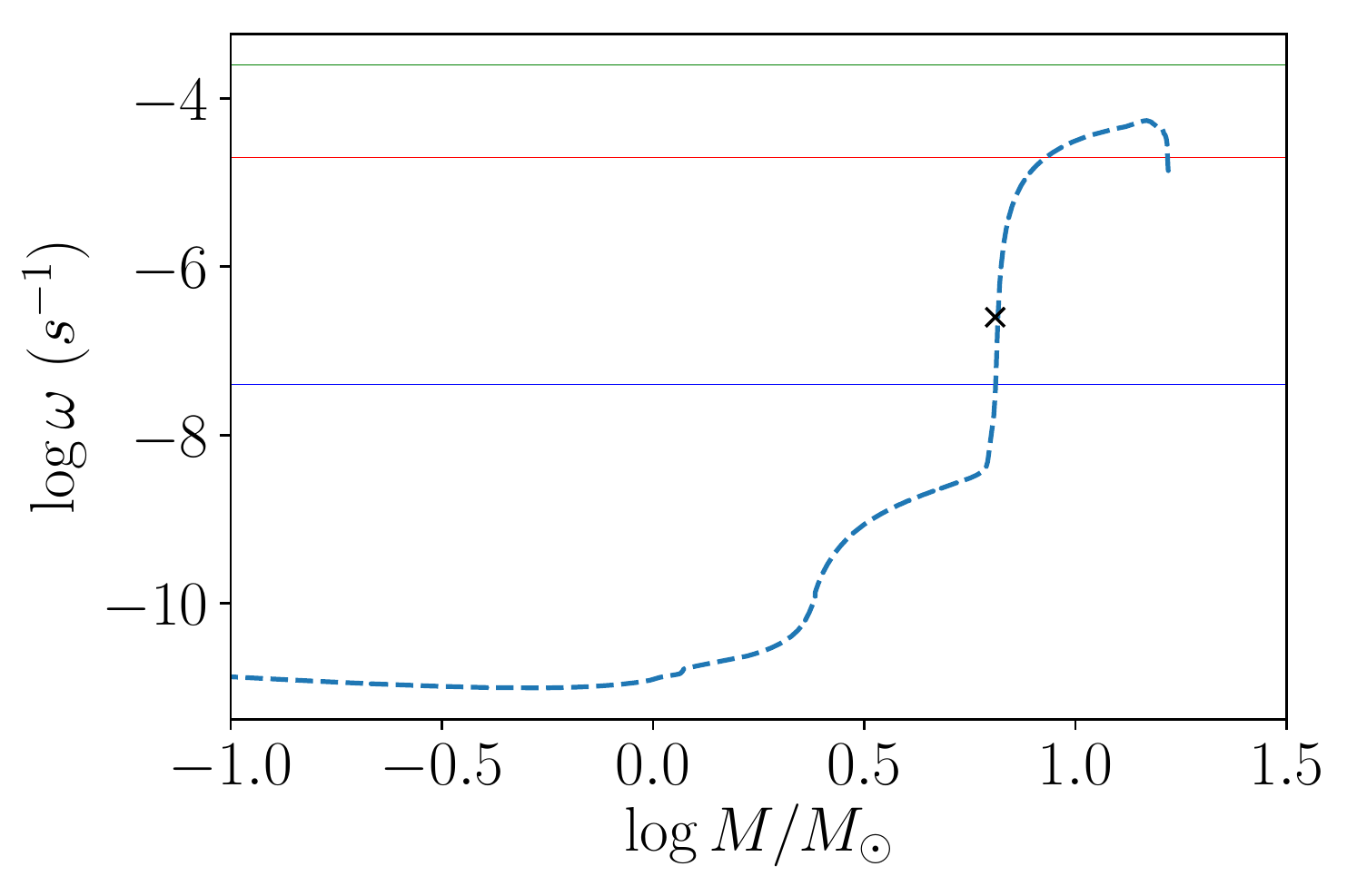}}
\caption{The wave propagation cutoff frequency given by equation \ref{cut} (dashed line) is 
given as a function of mass for the non-rotating 20~\m\ model in panel a at the minimum in luminosity 
at the base of the RSB (Figure \ref{trhohelium}) and in panel b in shell carbon burning 
(Figure \ref{trhocarbon}). The X marks the cutoff frequency evaluated at half
the radius of the star. In panel a, the blue line at $log~\omega \sim -7.3$ gives the frequency 
of the outer convective envelope, the red line at $log~\omega \sim -6.7$ gives the frequency 
generated by the inner convective hydrogen shell, and the green line at $log~\omega \sim -5.5$
gives the frequency generated in the helium core (see Figure \ref{Frequencies_Lmin}). 
In panel b, the blue line at $log~\omega \sim -7.5$ gives the frequency of the outer convective 
envelope, the red line at $log~\omega \sim -4.7$ gives the frequency generated by the 
helium shell, and the green line at $log~\omega \sim -3.7$ gives the frequency generated 
in the carbon shell (see Figure \ref{Frequencies_Cburn_nonrot}). 
}
\label{cutoff}
\end{figure}

We must also consider the cutoff frequency in the outer envelope. If the
frequency of a wave generated within the core is too low, that wave will not be 
able to propagate through the envelope. \citet{SQ14} estimated the cut--off 
frequency of the outer convective envelope as
\begin{equation}
\omega_{cut} \approx \frac{c_{s}}{H_p}
\label{cut}
\end{equation}
where $c_{s}$ is the sound speed and $H_p$ the pressure scale height in the envelope 
measured at 1/2 of the radius of the model. In order to propagate to the surface, 
$\omega_{conv} > \omega_{cut}$. While this prescription gives a useful
guideline, we note that the cutoff frequency can be a rather sensitive function of 
position in the envelope. Figure \ref{cutoff} gives the distribution of $\omega_{cut}$
in the envelope of the non-rotating 20~\m\ model at the luminosity minimum and 
in the stage of carbon shell burning; the rotating models are virtually indistinguishable
in this regard. In this figure, the X marks the frequency of $\omega_{cut}$ at the
half-radius point. Horizontal lines correspond to the frequencies generated by various 
features, as presented below. The low frequencies generated in the outer envelope
or the isolated H convective shell will not propagate, though large convective plumes
may reach the surface. Frequencies generated by the helium-burning core or the later 
convective helium shell may not propagate. The frequencies estimated for the carbon-burning
shell are expected to propagate through the envelope by this criterion. In the analysis
below, we present the frequencies generated by these various features as a function
of the mass of the stellar model and display them in comparison with $\omega_{cut}$
as computed by equation \ref{cut} at the half-radius point. The qualitative results for the 
lowest and highest frequencies are rather clear; the issue of the propagation of waves 
produced in the helium convective regions may be marginal. For the current discussion 
we ignore issues of wave dissipation; we return to this issue in \S \ref{detect}.

\begin{figure}%[htp]
\center
\includegraphics[width=3 in, angle=0]{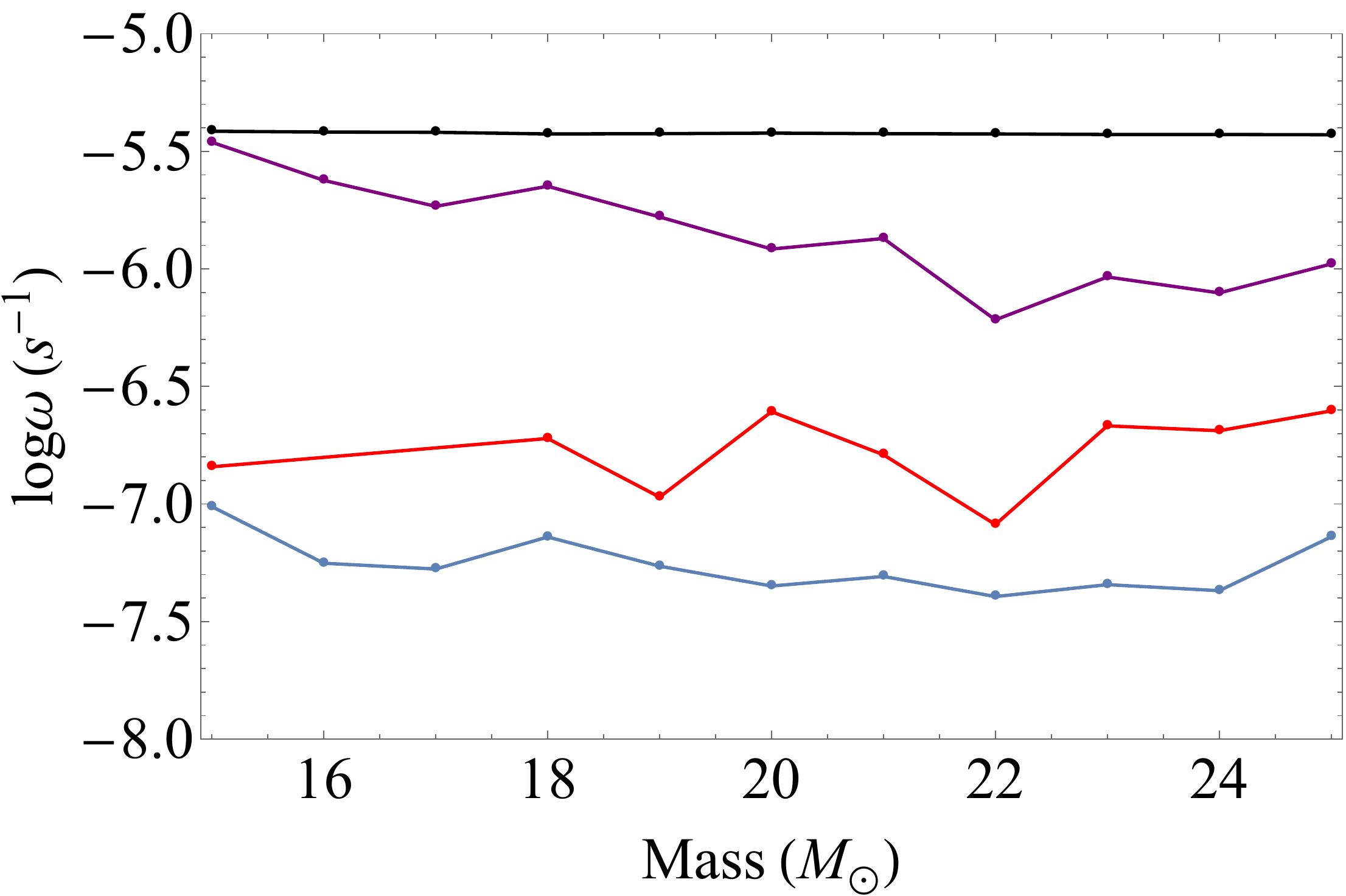}
\caption
{The distribution of convective frequencies and the half-radius envelope cut--off frequency as a 
function of ZAMS mass at the point of minimum luminosity at the base of the RSB for 
the non--rotating models. Similar convective structures are connected by the solid
lines. The purple curve around $log~\omega \sim -6$ denotes the cut-off frequency, 
below which none of the characteristic convective frequencies are expected to propagate 
to the surface. The blue curve around $log~\omega \sim -7.3$ corresponds to the outer 
convective envelope with a characteristic period, $T_{conv}$, of years. The red curve 
around $log~\omega \sim -7$ corresponds to a separate convective shell with 
$T \sim 10^6$ K that has a period about 3 times less. Only the convection in the 
helium core with a frequency $log~\omega \sim -5.5$ denoted by the black line 
might be observable at the surface with an overturn time of $\sim$ 20 d.
\label{Frequencies_Lmin}}
\end{figure}

Figure \ref{Frequencies_Lmin} gives the distribution of convective frequencies 
and the half-radius envelope cut-off frequency as a function of ZAMS mass at the point of 
minimum luminosity at the base of the RSB for the non--rotating models. 
These models are early in core helium burning. The outer convective envelope 
shows the characteristic timescale, $T \sim$ years, that is characteristic of the 
420 day pulsation and slow overturn of the convective plumes. As illustrated in Figures 
\ref{trhohelium} and \ref{trhocarbon}, models of the type we are studying often have 
a deeper hydrogen convective shell at $T \sim 10^6$ K that is separated from the outer 
convective envelope by a radiative region. This deeper convective hydrogen--rich 
region has a  characteristic timescale of $T \sim$ months. The timescale 
corresponding to the cut--off frequency varies from 20 to 70 days. Only the inner 
helium--burning convective core with a characteristic time of $T_{conv} \sim 20$~d 
might have the possibility to generate pressure waves that could propagate to the 
surface. The frequency of this mode varies by about 20\% over the range of our models, 
too small to be seen easily on this log plot.

\begin{figure}%[htp]
\center
\includegraphics[width=3 in, angle=0]{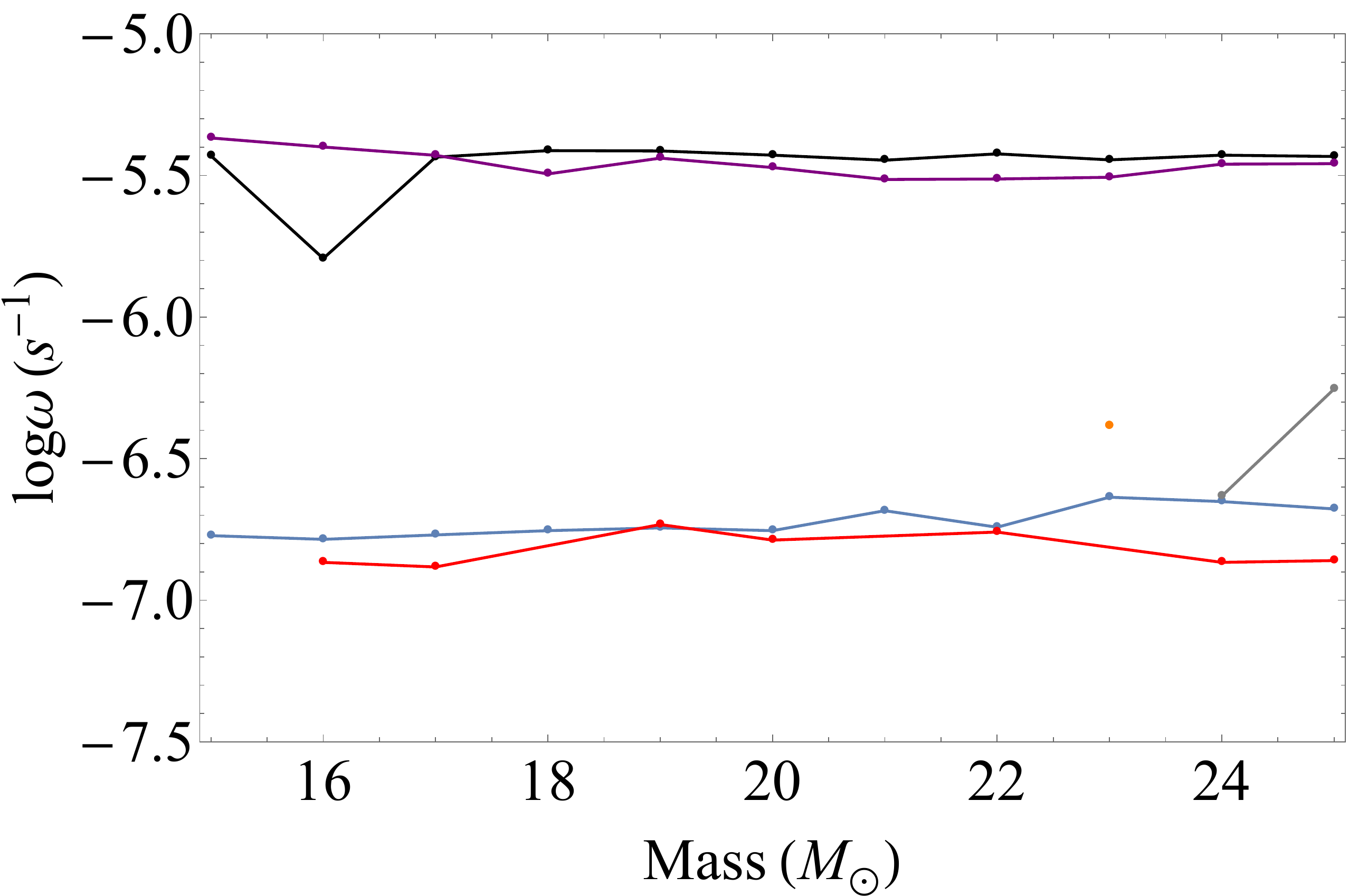}
\caption
{Similar to Figure \ref{Frequencies_Lmin} for the rotating models at the base 
of the RSB where they have rotational velocity $\sim 15$~\kms. The purple curve 
around $log~\omega \sim -5.5$ denotes the cut-off frequency. The blue curve around 
$log~\omega \sim -6.8$ corresponds to the outer convective envelope with a 
characteristic period of years. The red curve around $log~\omega \sim -6.8$ 
corresponds to the convective hydrogen shell at $T \sim 10^6$ K. Even the 
convection in the helium core with a frequency $log~\omega \sim -5.4$ and
a period of $\sim$ 20~d denoted by the black line might not be observable 
at the surface.
\label{Frequencies_v15}}
\end{figure}

Figure \ref{Frequencies_v15} gives the distribution of convective frequencies and the 
half-radius envelope cut-off frequency as a function of ZAMS mass at the point near where the 
rotational velocity is $\sim 15$~\kms\ near the base of the RSB for rotating models 
\citep{Wheeler17}. Figures \ref{trhohelium} and \ref{trhocarbon} show that while
there is some difference in the structure of the outer envelope in the rotating model,
there is relatively little difference in the deeper interior. The outer convective envelope 
of the rotating model shows a characteristic period of about a year, similar to that in 
Figure \ref{Frequencies_Lmin}, but somewhat shorter since the evolution is at a 
slightly earlier phase. The deeper convective hydrogen shell at $T \sim 10^6$ K has 
nearly the same period as in Figure \ref{Frequencies_Lmin} and very similar to that 
of the outer envelope. The inner helium--burning convective core again has a 
characteristic period of $\sim 20$~d, but so too does the cut-off frequency that 
does not show the increase in period with mass that characterized the cut-off frequency
in Figure \ref{Frequencies_Lmin}. Taken at face value, the models that formally 
match the observed rotation velocity might be expected to reveal no inner signals. 
In principle, the detection of such signals with a period of about 20 days might 
allow discrimination of non-rotating from rotating models early in the core 
helium--burning phase.  

\begin{figure}%[htp]
\center
\includegraphics[width=3 in, angle=0]{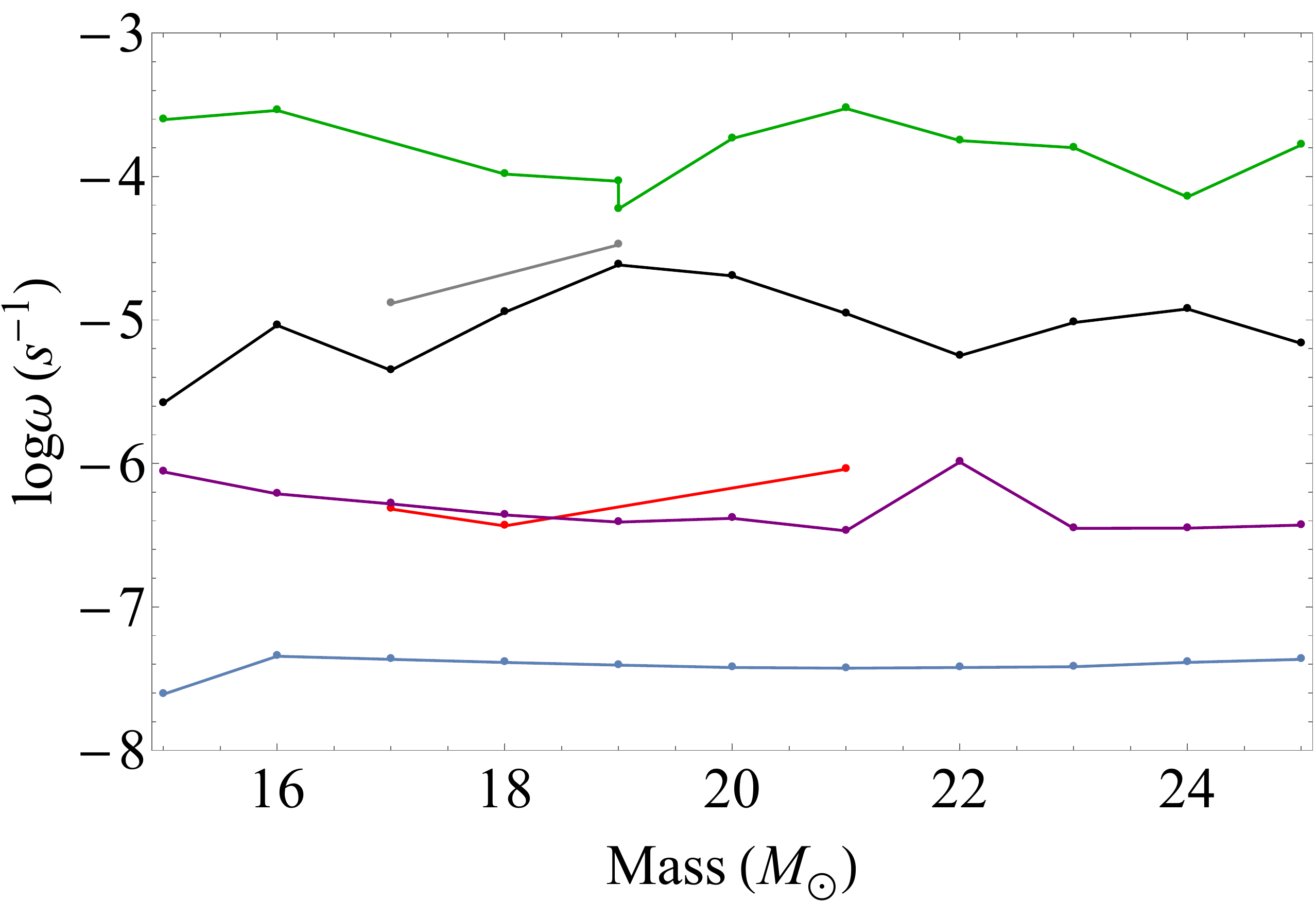}
\caption
{Similar to Figure \ref{Frequencies_Lmin} for carbon shell burning for the 
non--rotating models. The purple curve around $log~\omega \sim -6.2$ denotes 
the cut-off frequency. The blue curve around $log~\omega \sim -7.2$ corresponds 
to the outer convective envelope with a characteristic period of years. The red 
curve around $log~\omega \sim -6$ corresponds to the separate convective hydrogen 
shell with $T\sim10^6$ K. The black curve around $log~\omega \sim -5$ corresponds 
to the helium--burning convective shell. The green curve around $log~\omega \sim -4$ 
corresponds to the carbon--burning shell. Intermediate convective structures are 
formed at some masses (gray curve). The convection in the deeper convective 
helium- and carbon-burning shells might generate gravity and pressure waves that 
propagate to the surface with periods of about 10~d and 1d, respectively.
\label{Frequencies_Cburn_nonrot}}
\end{figure}

Figure \ref{Frequencies_Cburn_nonrot} gives the distribution of convective 
frequencies and the half-radius envelope cut--off frequency as a function of ZAMS mass 
at the carbon shell-burning phase for the non--rotating models. Here the 
structure of the convective regions is more complex. We have attempted to 
identify similar regions in each ZAMS mass but note that there are some models 
that do not display a particular structure when adjacent masses do (hence 
a missing ``dot" in the figure) and others where a given structure is displayed 
by only a limited mass range. We have only attempted to capture regions of 
full convection and have neglected regions of semi--convection on the grounds 
that their slower overturn will lead to low frequencies and low power. The 
outer envelope again shows a period of years. The timescale corresponding
to the cut--off frequency is about 70 days. The characteristic period 
of the separate hydrogen--rich convective shell with $T \sim 10^6$ K again has
a period of months, is quite close to the cutoff frequency for many models, and
is manifest for a restricted set of models, 17, 18, and 21. A principle feature 
of these models is the convective helium--burning shell with a period at about 
10~d and the convective carbon--burning shell at about 1~d or less (both with 
significant variation with mass) that could, in principle, produce waves that propagate 
to the surface, indicating the presence of those structures. Model 19 has two 
inner convective regions with nearly the same frequency, one at $m_r = 1.30$ \msun\ 
and one at slightly higher frequency at $m_r = 1.13$ \msun.  Models 17 and 19 
have an intermediate convective layer at about $m_r = 2.5$ \msun, somewhat 
deeper than the helium convective shell and with somewhat higher frequency. 

\begin{figure}%[htp]
\center
\includegraphics[width=3 in, angle=0]{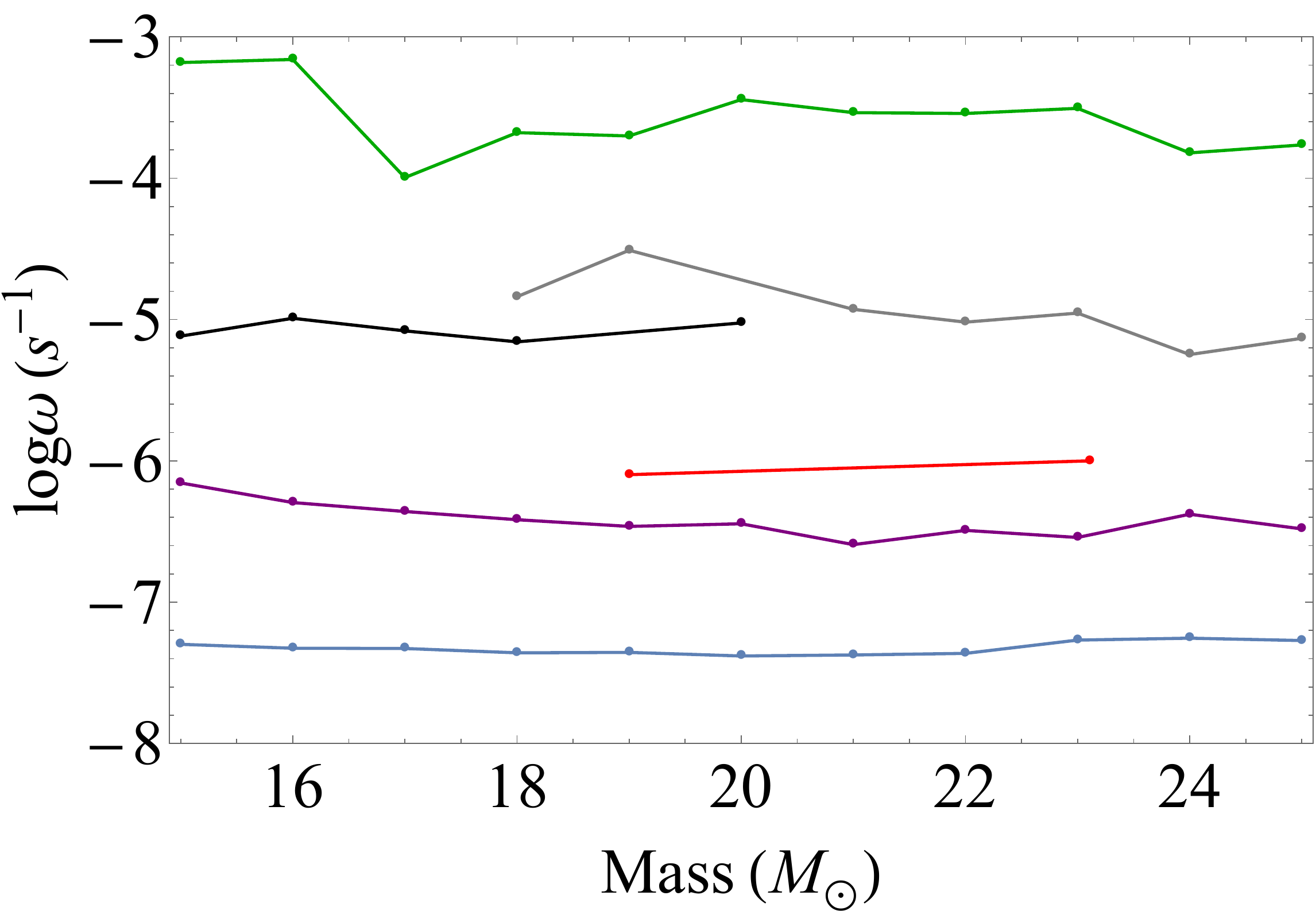}
\caption
{Similar to Figure \ref{Frequencies_Cburn_nonrot} at carbon shell
burning for the rotating models. The purple curve around $log~\omega \sim -6.2$ 
denotes the cut-off frequency. The blue curve around $log~\omega \sim -7.2$ 
corresponds to the outer convective envelope with a characteristic period of 
years. The red points correspond to the convective hydrogen shell at $T \sim 10^6$~K. 
The black and gray curves around $log~\omega \sim -5$ correspond to the convective 
helium--burning shell. The green curve at $log~\omega \sim -3$ to -4 corresponds 
to the carbon--burning shell. The convection in the convective helium-- and 
carbon--burning shells might generate gravity and pressure waves that propagate to 
the surface with periods of about 10~d and 0.1 to 1~d, respectively. The intermediate 
hydrogen convective structures might also have potentially observable frequencies.
\label{Frequencies_Cburn_rot}}
\end{figure}

Figure \ref{Frequencies_Cburn_rot} gives the distribution of convective 
frequencies and the half-radius envelope cut--off frequency as a function of ZAMS mass 
at the carbon shell-burning phase for the rotating models. The outer envelope 
shows the characteristic period of years. The hydrogen shell at $T = 10^6$ K
appears only sporadically in the models but with characteristic period of
months. The characteristic period of the convective helium shell is again about 
10~d. The inner carbon shell has a period of a few tenths of a day. With the 
exception of the outer convective envelope, all the convective regions illustrated 
could, in principle, produce waves that propagate to the surface. As for the 
non--rotating models, the structure near carbon burning is expected to produce 
a more complex spectrum with generally higher frequencies than the models 
near the base of the giant branch.

\section{Detectability}
\label{detect}

Generating gravity waves in convective regions is one thing, getting a signal
to the surface is quite another. There are several stages in the process
as discussed by \citet{QS12}, \citet{SQ14} and \citet{Fuller17}. Only a 
fraction of the convective energy is converted to gravity waves that 
propagate in radiative regions where their frequency is less than the 
Brunt-Vaisala frequency. These waves must then pass through outer
convective regions where they evanesce with a certain damping rate
\citep{Unno89}. Note that these outer convective regions can generate 
their own gravity waves. Eventually, the gravity waves will reach a position 
in the star at which their frequency exceeds the Lamb frequency, 
$\sqrt{\ell(\ell +1)} c_s/r$, where $c_s$ is the local sound speed and $\ell$ 
is the mode order. At that point, the waves will be converted to acoustic 
waves. In practice, this conversion occurs around the outer boundary
of the helium core where the density and sound speed drop significantly.
These acoustic waves will strengthen as they propagate down the density 
gradient into the outer envelope. If they develop into shocks or lose their 
energy to thermal diffusion, they will be dissipated and not reach the surface. 
There are a variety of other effects that must be taken into account, such 
as neutrino damping in the core, and wave reflections from composition 
boundaries in which the density gradient is steep and hence the WKB
approximation breaks down. 

Here we will address a few of the major factors, especially the efficiency 
of converting convective power into gravity waves and the thermal 
dissipation in the envelope. Following \citet{SQ14} and \citet{Fuller17}, 
we will only consider modes of angular frequency given by equation 
\ref{char_freq} and with $\ell = 1$ since lower frequency waves tend to 
be more highly damped by shocks (but see the discussion on thermal 
diffusion which is less for low frequency; equation \ref{damp}) and higher 
frequencies from a given source contain less power, and waves with larger 
values of $\ell$ both contain less power and are more strongly damped. 
\citet{Fuller17} finds that the loss of power due to evanescence is a 
factor between 2 and 10. This is a significant factor, but small compared 
to other uncertainties. We omit this factor from our formal calculations, 
but note its role and return to a discussion of this factor in \S\ref{thermal}. 
The acoustic waves that might propagate to the surface 
have frequencies much higher than the envelope pulsation mode of 420 days 
or the longer envelope convective overturn time. These structures are 
effectively ``frozen" on the timescales of the inner convective regions.

In the following discussion we focus on the non-rotating model with
ZAMS mass of 20 \msun. This mass is the most likely value for
Betelgeuse (but not, of course, for other RSG), and we find that 
rotation does not substantially affect the issues involved.

\subsection{Convective Power}
\label{power}

To estimate the possible detectability of the characteristic frequency from the 
convective noise, we first estimated the power associated with convective 
kinetic energy in each convective region as
\begin{equation}
L_{conv} \sim \frac{1}{2} \Delta M_{conv} v_{conv}^2 T_{conv}^{-1} \sim 
 \frac{1}{4\pi} \Delta M_{conv} \frac{v_{conv}^3}{H_p}
\label{lum1} 
\end{equation}
where $\Delta M_{conv}$ is the mass in the convective region. Following 
\citet{coxgiuli} (\S\S14.2, 14.3, equations 14.113 and 14.114), a more 
rigorous estimate allowing for radiative losses yields
\begin{equation}
L_{conv} \approx 4 \frac{dm}{dr}v_{conv}^3,
\label{lum2} 
\end{equation}
which agrees with equation \ref{lum1} within factors of a few. 

\begin{figure}%[htp]
\center
\includegraphics[width=3 in, angle=0]{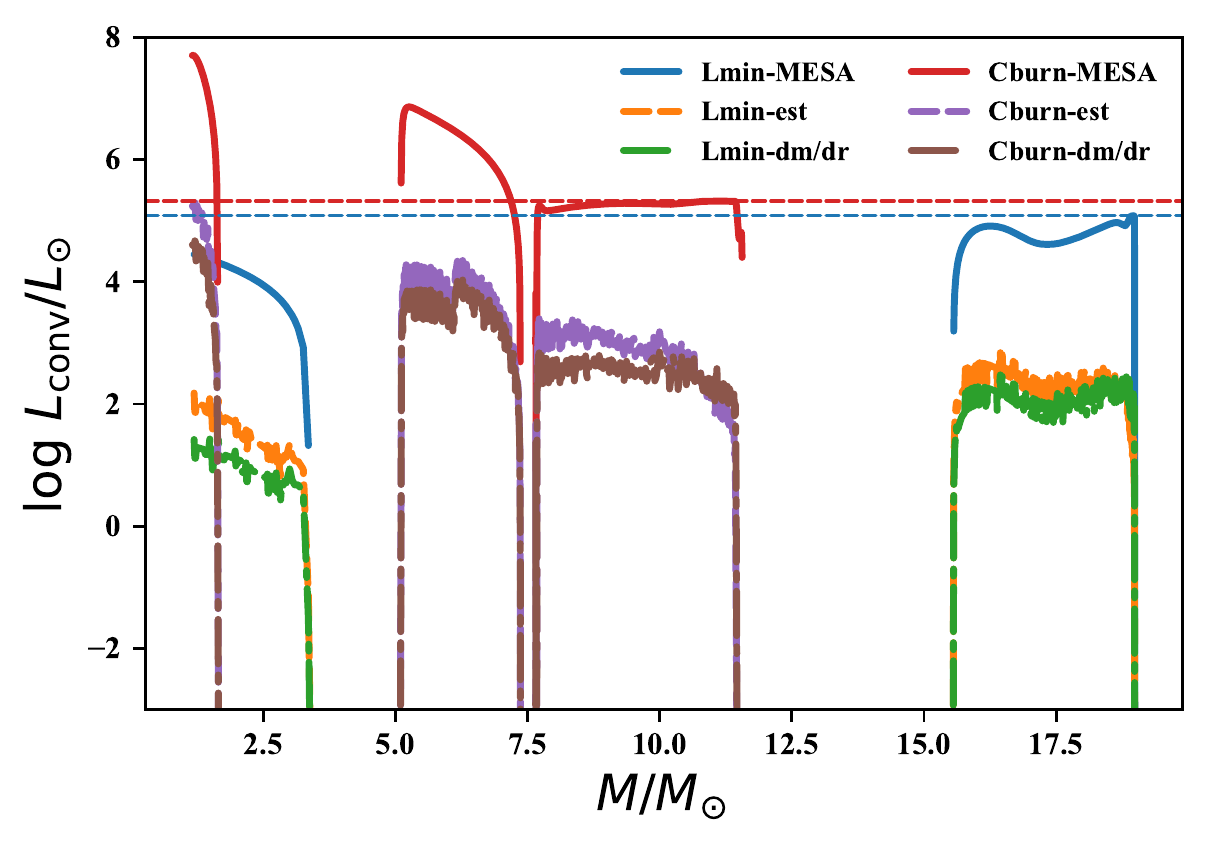}
\caption
{Three estimates of the convective luminosity are given as a function
of mass for the 20 \msun\ non-rotating model. The solid lines present
the parameter $L_{conv}$ computed by $MESA$ using mixing-length
theory; blue represents core helium burning near the base of the RSB, 
red the phase of shell carbon burning. The lower estimates that roughly
agree with one another given by the dashed lines are based on 
equations \ref{lum1} (orange and purple for core helium burning and
carbon shell burning, respectively) and \ref{lum2} (green and brown 
for core helium burning and carbon shell burning, respectively) employing 
the value of $v_{conv}$, $H_P$ and $dm/dr$ computed by $MESA$.
The horizontal dashed lines (blue for the base of the RSB and red for
carbon shell burning) indicate the level of the total luminosity radiated 
at the photosphere.
\label{lconv1}}
\end{figure}

Interestingly, for the RSG models we study here, these prescriptions
give a value of $L_{conv}$ that is smaller than the value computed
directly by $MESA$ by a factor $\sim 100$. The estimates in equations
\ref{lum1} and \ref{lum1} assumed that a factor $c_p \rho T/P$ was
of order unity. Closer inspection revealed that not to be the case. 
This factor averages of order 10 to 20 both at the base of the RSB
and in carbon shell burning and spikes to close to 50 at certain
locations in the latter case. Figure \ref{lconv1} shows the convective 
luminosity as a function of mass estimated by the three methods, 
equation \ref{lum1} and equation \ref{lum2}, using the values of $v_{conv}$, 
$H_P$, and $dm/dr$ computed by $MESA$, and the $MESA$ value for 
$L_{conv}$ in core helium burning at the base of the RSB and in 
carbon shell burning for the model of 20 \msun. For the subsequent
discussion, we employ the $MESA$ value for $L_{conv}$. Figure \ref{lconv1} 
also gives the total luminosity at the photosphere for the model at the 
base of the RSB and in carbon shell burning. Note that the convective 
luminosity in the helium-burning core is less than the radiated luminosity 
while that in the helium-burning shell is comparable to the radiated 
luminosity. The convective luminosity in the carbon-burning shell 
substantially exceeds the radiated luminosity. For illustration, we 
take the value of the convective luminosity at the radial midpoint in 
each region in the following discussion. 

Table \ref{tab:power} gives the period, $T_{conv} = 2 \pi /\omega_{conv}$, and
convective luminosity for a representative sample of convective zones 
for the non-rotating 20~\msun\ model in order of increasing depth in the model. Only modes 
with frequencies comparable to or above the envelope cut--off frequency are presented.
\citet{GK90} argued that the efficiency of conversion of convective energy into
the energy of gravity waves is approximately $\mathcal{M}$, the mean Mach
number in the convective region. \citet{LQ13} estimate that the efficiency might
be higher, scaling as $\mathcal{M}^{5/8}$. Table \ref{tab:power} also gives estimated 
power in gravity waves for this model using these two efficiency factors.

Taking the results of Table \ref{tab:power} at face value, the inner convection could
deliver appreciable power to the surface of the star. The helium core could produce
$\sim 10^{36}$ \ergs\ and the convective carbon shell phase up to $\sim 10^{38}$ \ergs,
especially if the convection is efficient in producing acoustic waves. Such
changes might produce interesting effects. At this point, we have not yet accounted for 
various efficiency and dissipation factors. More important, to be significant in an
asteroseismic context, the critical factor is not an increment in power radiated, but
whether the acoustic waves can perturb the surface luminosity in a manner that
carries information about the deep convective layers. 

%\begin{landscape}
\begin{table}
\caption{Estimated Gravity-Wave Power Generated at the Radial Midpoint in Convective Regions 
for the Non-rotating 20~\msun\ Model.}
\label{tab:power}
\begin{tabular}{lccccc}
\hline
& $T_{conv}$ & $L_{conv}$ & $\mathcal{M}$ & $L_{conv}$$\mathcal{M}$ & $L_{conv}$$\mathcal{M}^{5/8}$  \\
& days & $10^{36}$     & $10^{-4}$         & $10^{36}$    & $10^{36}$ \\
&         &  erg s$^{-1}$ &  & erg s$^{-1}$ &  erg s$^{-1}$  \\

\hline
$L_{min}$ &     &     &     &     &    \\
\\
He core        & 21.1 & 133 & 2.04 & 0.0271 & 0.656 \\
                     &     &     &     &     &     \\                                                     
{\sl C burn}   &     &     &     &     &     \\    
\\                                                                    
He shell   &  13.8 & 3470 & 24.6 &  8.55 & 81.4 \\
              &     &     &     &     &     \\       
C shell   &  0.14  & 7620   &  1.25   &  0.953   & 27.7 \\
              &     &     &     &     &      \\         
\hline          

%\footnote{determined at the radial midpoint of the convective region}    
              
\end{tabular}
\end{table}
%\end{landscape}

\subsection{Wave Amplitudes}
\label{amplitude}

The acoustic waves gain in amplitude as they run down density gradients, 
especially the sharp gradient that marks the passage from the outer helium 
core to the hydrogen envelope. Defining $\xi$ to be the amplitude or radial
displacement of the wave, the luminosity of the wave can be written
\begin{equation}
L_{wave} = 4 \pi r^2 \rho (\xi \omega)^2 c_s
\label{lwave}
\end{equation}
so that with wavenumber $k = \omega/c_s$ the dimensionless quantity
$k\xi$, the displacement relative to the wavelength, can be written
\begin{equation}
k \xi = \left(\frac{L_{wave}}{4\pi r^2 \rho c_s^3}\right)^{1/2} \approx 
8.9\times10^{-5} \left(\frac{L_{wave,36}}{r_{11}^2 \rho c_{s,7}^3}\right)^{1/2}
\label{kxi}
\end{equation}
where $L_{wave,36}$ is the wave amplitude in units of $10^{36}$ \ergs,
a typical luminosity for the weak waves we consider here (Table \ref{tab:power}), 
$r_{11}$ is the radius in units of $10^{11}$ cm that is characteristic of the helium
core, and $c_{s,7}$ is the sound speed in units of $10^7$ \cms. \citet{QS12} 
give a similar expression. This shows that the initial amplitude of the waves
is small. 

Figure \ref{amplitude} shows the relative displacement, $k \xi$, versus radius 
for waves generated in the convective helium core when near the base of the 
RSB and in the convective helium and carbon shells when in the carbon 
shell-burning phase for the non-rotating 20~\msun\  model. In the outer part of
the helium core, $r_{11} \sim 4$, the density is near 1 \gcm3, and the sound 
speed is $c_{s,7} \sim 10$. The value of $k \xi$ is of order $10^{-6}$ for all 
models. From this low value, the outer edge of the model corresponding to 
the base of the RBG is reached before $k \xi$ reaches unity. In
the carbon shell-burning phase, $k \xi$ exceeds unity in the outer reaches
of the model.

\begin{figure}%[htp]
\center
\includegraphics[width=3 in, angle=0]{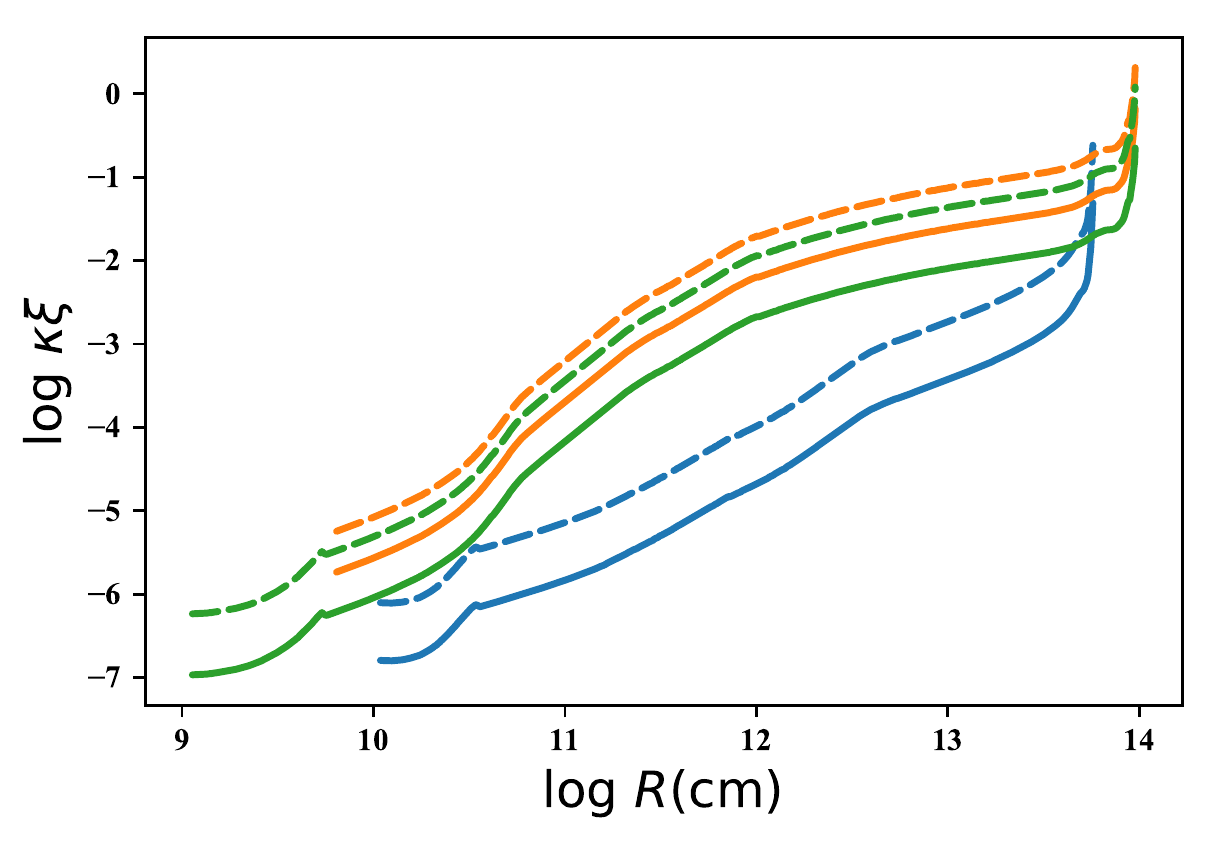}
\caption
{The relative displacement, $k \xi$, (equation \ref{kxi}) is given as
a function of radius for gravity waves emitted from the convective
helium-burning core at the base of the RSB (blue) and from the 
carbon (green) and helium (orange) convective burning shells in 
carbon shell burning for the non-rotating 20~\msun\ model. 
The corresponding gravity-wave luminosities in each case
are given in Table \ref{tab:power}; values of $k \xi$ corresponding 
to the convective luminosity reduced by the Mach number, 
$L_{conv}$$\mathcal{M}$, are given by the solid lines and the 
dashed lines correspond to $L_{conv}$$\mathcal{M}^{5/8}$.
\label{amplitude}}
\end{figure}

\subsection{Modulation of Surface Luminosity}
\label{modulation}
 
The modulation of the luminosity due to linear adiabatic modulation by acoustic 
waves is very roughly given by \citep{Unno89}
\begin{equation}
\frac{|\Delta L|}{L} \sim k \xi.
\label{deltaL}
\end{equation}
The perturbation of the luminosity radiated at the surface due to perturbations in the 
temperature, density, and surface area can be approximated by evaluating
equation \ref{kxi} at the location in the envelope where $\omega \tau_{therm} \sim 1$,
where $\tau_{therm}$ is the local thermal timescale,
\begin{equation}
\tau_{therm} = \frac{4 \pi r^2 H_p \rho c_P T}{L} 
\label{therm}
\end{equation}
and where $c_P$ is the specific heat at constant pressure and $L$ is the radiated flux 
\citep{pfahl08, Fuller17heartbeat}. 

%Combining Equations \ref{kxi} and \ref{therm}, 
%we can formally write
%\begin{equation}
%\frac{|\Delta L|_{surf}}{L} \sim \left(\frac{4 \pi L_{wave}}{\rho c_s}\right)^{1/2} \frac{H_p c_P T}{L} 
%\label{perturb}
%\end{equation}
%where the LHS is the relative modulation of the surface luminosity, and the 
%RHS is evaluated at the radius where $\omega \tau_{therm} = 1$. 
We sought to evaluate equation \ref{deltaL} by computing 
$\omega \tau_{therm}$ as a function of $r$ for a given frequency of interest 
as given in Table \ref{tab:damping} and then 
%invoking Equation \ref{kxi}, evaluating $\xi/r$ as
%\begin{equation}
%\frac{|\Delta L|}{L} \sim \frac{|\xi|}{r} =  \left(\frac{L_{wave}}{4 \pi r^4 \rho c_s \omega^2}\right)^{1/2} 
%\label{perturbL}
%\end{equation}
evaluating equation \ref{kxi} at that radius. The perturbation to the luminosity 
in magnitudes due to the modulation by the convective noise would then be
\begin{equation}
\delta m = \left| -2.5 {\rm log} \left(\frac{L_{rad} + |\Delta L_{surf}|}{L_{rad}}\right)\right| \approx 1.1 \frac{|\Delta L_{surf}|}{L_{rad}}
\label{perturbM}
\end{equation}
where $|\Delta L_{surf}| << L_{rad}$ is the absolute value of our estimate of the 
observable perturbation of the luminosity at the surface.

\begin{table}
\caption{Frequencies$^1$ Employed to Compute Wave Amplitudes and Damping Masses}
\label{tab:damping}
\begin{tabular}{lccc}
\hline
Model Mass & Helium Core & Helium Shell & Carbon Shell \\

\hline

15 & $3.5\times10^{-6}$ & $5.3\times10^{-6}$ & $5.2\times10^{-4}$ \\

20 & $3.8\times10^{-6}$ & $2.0\times10^{-5}$ & $1.8\times10^{-4}$ \\

25 & $3.7\times10^{-6}$ & $6.9\times10^{-6}$ & $1.7\times10^{-4}$ \\

\hline    
$^1$ in units of $s^{-1}$
                             
\end{tabular}
\end{table}

In practice, we found this prescription difficult to implement in a 
straightforward manner. For the non-rotating 20 \msun\ model
at the base of the RSB, we found that $\omega \tau_{therm} = 1$ 
only very close to the surface. For the model in carbon shell burning,
$\omega \tau_{therm}$ did not reach unity with the resolution of 
the current model. The implication is that inasmuch as $\omega \tau_{therm}$
nears unity, it is at a radius that is essentially equal to the outer radius of 
the model. We can then evaluate Equation \ref{deltaL} at the outer
radius of the model. For the models we consider here, typical values
for the density and sound speed in the outer layers are $10^{-8}$ \gcm3\ 
and $10^6$ cm $s^{-1}$ so we can estimate
\begin{equation}
\frac{|\Delta L_{surf}|}{L} \sim 2.8\times 10^{-2}  \left(\frac{L_{wave,36}}{R_{14}^2 \rho_{-8} c_{s,6}^3}\right)^{1/2} 
\label{perturbL}
\end{equation}

For the non-rotating 20~\msun\ model, the ratio of the amplitude of the 
perturbed luminosity to the total luminosity (neglecting any subsequent damping 
between the source and the surface for now) estimated in this way is given in 
Table \ref{tab:delta} (the parameter $d_a$ is discussed below, equation \ref{perturb}), 
assuming all the factors in the denominator of equation \ref{perturbL} are unity. 
The estimated variation in magnitudes would be larger than $\Delta L_{surf}/L$ by 
about 10\%, according to equation \ref{perturbM}. At this point, the results are encouraging, 
ranging from potential perturbations of order several millimagnitudes up to a few 
tenths of a magnitude, well within the range of astronomical detection.

\begin{table}
\caption{Estimated Perturbation of the Surface Luminosity, $\frac{|\Delta L_{surf}|}{L}$,
by Acoustic Waves for the Non-rotating 20~\msun\ Model.}
\label{tab:delta}
\begin{center}
\begin{tabular}{lccc}
\hline
&  $L_{conv}$$\mathcal{M}$ & $L_{conv}$$\mathcal{M}^{5/8}$ & ${\rm log}~d_a$  \\ 
\hline
%&  $\frac{|\Delta L|_{surf}}{L}$  & $\delta m$ & $\frac{|\Delta L|_{surf}}{L}$ & $\delta m$ \\
%&     & millimag         &     & millimag \\
%\hline
$L_{min}$ &     &   &   \\
\\
He core        & 0.0047 &  0.023 & 1.6  \\
                     &     &   & \\                                                     
{\sl C burn}   &     &  &  \\    
\\                                                                    
He shell   &  0.083 & 0.26 & 220 \\
              &     &    &   \\       
C shell   &  0.028  &   0.15 & 22,000  \\
              &     &   &    \\         
\hline

\end{tabular}
\end{center}  
\end{table}

%\footnote{determined at the radial midpoint of the convective region}    

\subsection{Wave Dissipation Processes}
\label{dissipation}

Even if the wave powers estimated in Table \ref{tab:power} are appropriate,
the acoustic waves have yet another gauntlet to run before imparting any
perturbation to the surface luminosity as estimated in Table \ref{tab:delta}. 
They must avoid dissipation in the envelope either by growing to non-linear 
amplitudes and becoming shocks or losing their energy to thermal diffusion. 

\subsubsection{Shock dissipation}
\label{shock}

In our models, the density typically drops from $\rho \sim 10$ \gcm3\ in the outer 
helium envelope to $\rho \sim 10^{-6}$ \gcm3\ at the base of the hydrogen 
envelope. For the latter density, equation \ref{kxi} gives $k \xi \sim 0.1$, suggesting 
that these relatively weak waves may escape shock formation at that point. The 
density continues to decline in the hydrogen envelope, but the radius increases 
so there is a tradeoff that helps preserve the waves. 

Figure \ref{amplitude} gives the quantitative prediction of the profile of $k \xi$ 
for characteristic phases of the 20 \msun\ model. The relative displacement
remains less than $10^{-2}$ until the outermost layers. This suggests that
shocks will not necessarily form already at the base of the hydrogen envelope.
According to \citet{romatzner17}, shocks may form even more robustly
than we have estimated here by examining the behavior of $k \xi$.
This will exacerbate the likelihood that gravity waves from the carbon-burning
shell will dissipate in shocks at the base of the outer convective envelope,
but might still mean that the weak waves from core helium burning 
survive that process.

\subsubsection{Thermal Dissipation}
\label{thermal}

The other factor that imperils the propagation of the waves is the 
dissipation by thermal diffusion \citep{QS12,SQ14}. \citet{Fuller17}
expresses this as a damping mass
\begin{equation}
M_{damp} = \frac{3 \pi \rho^3 r^2 c_s^3 c_p \kappa}{4 \sigma T^3 \omega^2} \approx 
7\times10^5 M_{\rm \odot} \frac{\rho_{-6}^3 r_{12}^2 c_{s,7}^3}{T_4^3 \omega_{-6}^2},
\label{damp}
\end{equation}
where $c_p$ is the specific heat at constant pressure, 
$\kappa \sim 0.4$ cm$^{-2}$ g$^{-1}$ is the opacity, $\sigma$ is the 
Stefan-Boltzmann constant, $r_{12}$ is the radius in units of $10^{12}$ cm 
that is more characteristic of the hydrogen envelope, $\rho_{-6}$ is the
density in units of $10^{-6}$ g cm$^{-3}$, $T_4$ is the temperature in units 
of $10^4$ K, and $\omega_{-6}$ is the frequency in units of $10^{-6}$ s$^{-1}$.
This damping mass is an especially sensitive function of the density. If the
density plummets to a sufficiently low value in the hydrogen envelope, the
wave energy will be rapidly dissipated. The resulting heating can cause
the envelope to expand \citep{Fuller17} or even dynamically eject 
some of the envelope \citep{SQ14}. There is an implied threshold such
that once the waves begin to dissipate and expand the envelope, the
envelope density will decrease, increasing the tendency for the waves
to dissipate. 

\begin{figure}%[htp]
\center
\includegraphics[width=3 in, angle=0]{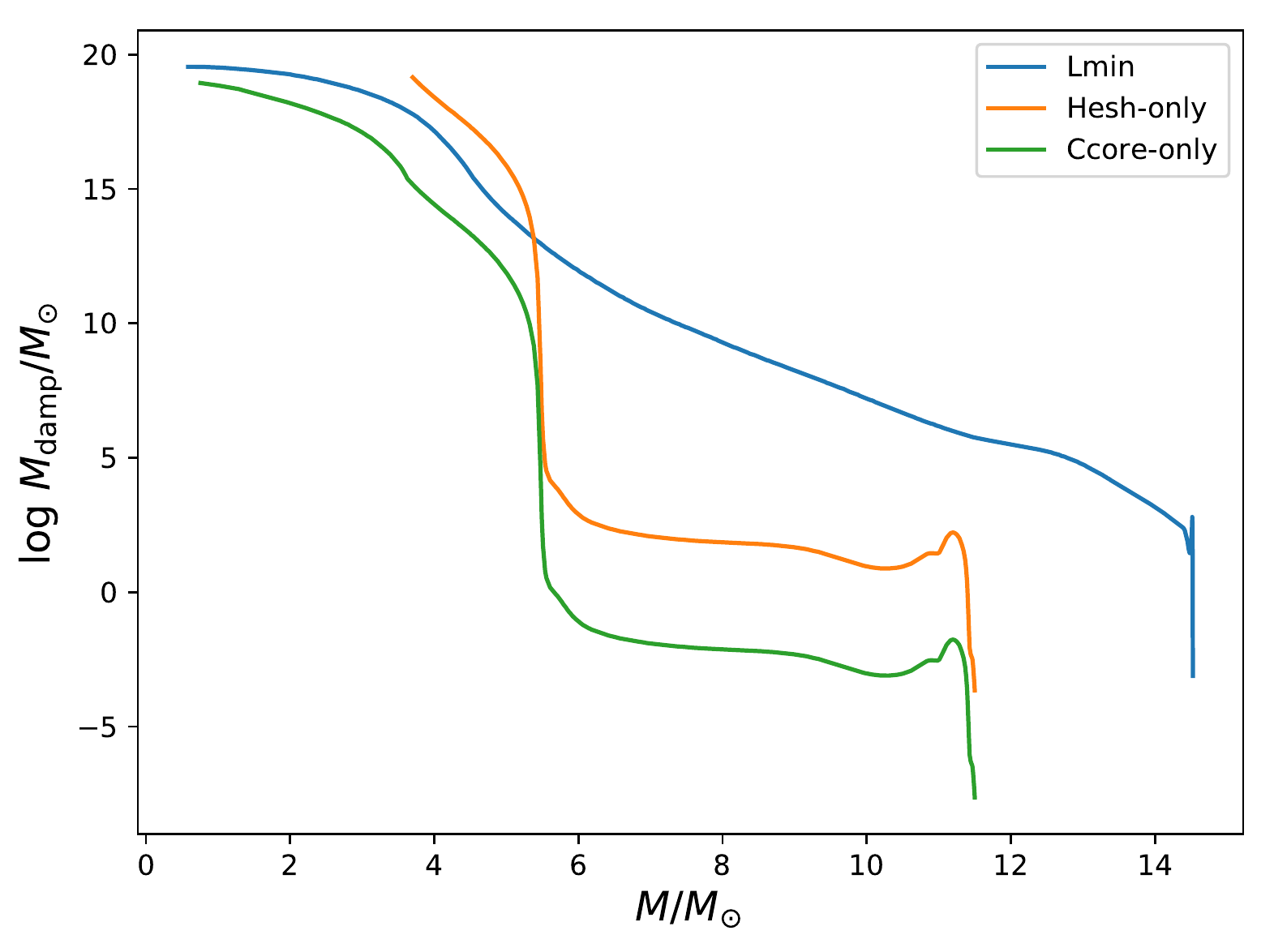}
\caption
{The distribution of the damping mass $M_{damp}$ from equation \ref{damp} 
at two evolutionary stages for the non-rotating model of 15 \msun. Blue 
corresponds to the model in convective core helium burning at the base
of the RGB, yellow to the helium burning shell during the first stage of
off-center convective carbon burning and green to that carbon-burning shell.
\label{damp15}}
\end{figure}

\begin{figure}%[htp]
\center
\includegraphics[width=3 in, angle=0]{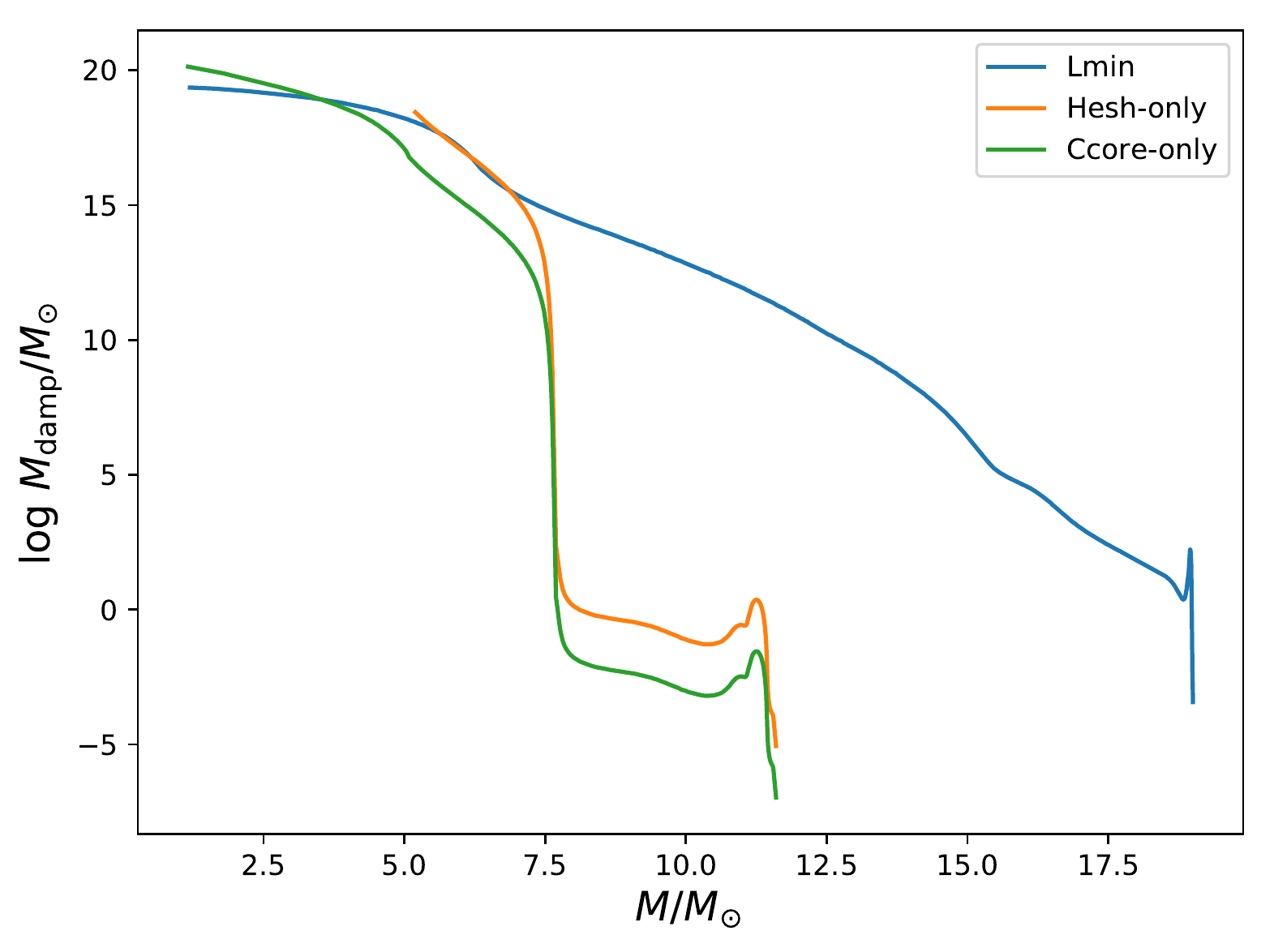}
\caption
{Similar to figure \ref{damp15}, but for the non-rotating model of 20 \msun.
\label{damp20}}
\end{figure}

\begin{figure}%[htp]
\center
\includegraphics[width=3 in, angle=0]{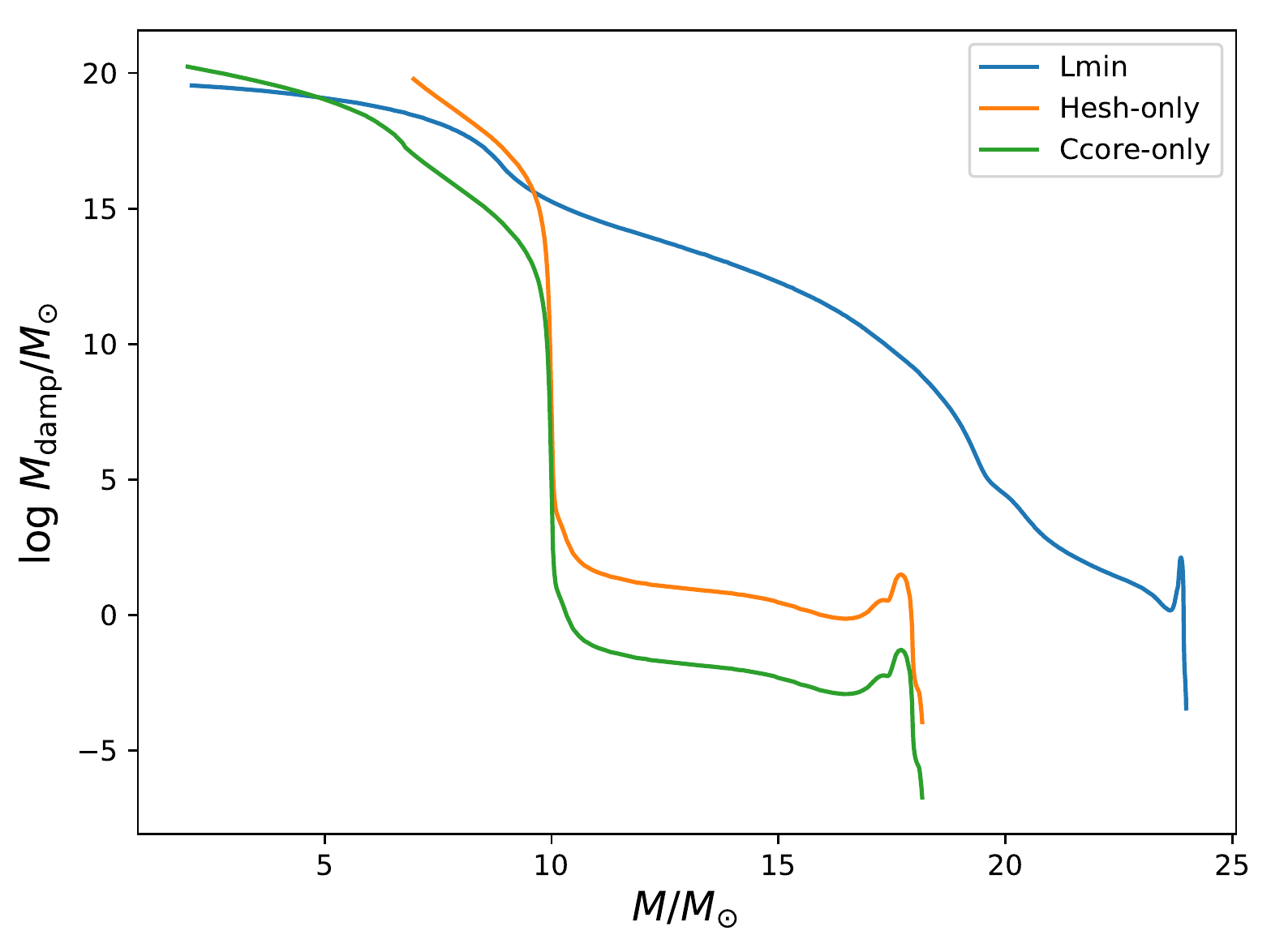}
\caption
{Similar to figure \ref{damp15}, but for the non-rotating model of 25 \msun.
\label{damp25}}
\end{figure}

The weak waves we consider here may or may not
dissipate. To check this, we have examined the density structure and hence
the variation of $M_{damp}$ for models in which the heating from
dissipation is ignored, leading to relatively large envelope densities.
Figures \ref{damp15} and \ref{damp20} give the distribution of the 
damping mass as a function of the mass of the model (that is subject to 
mass loss) at the given stage of evolution for our models of 15 and 
20 \msun, respectively, for core helium burning near the base of the RSB 
and in the stage of the first off-center convective shell carbon-burning
for non-rotating models. The corresponding frequencies are given in 
Table \ref{tab:damping}. We also computed a 25 \msun\ model, but had 
problems with model convergence with the default mass loss. Halving 
the ``Dutch" mass-loss rate from $\eta = 0.8$ to $\eta = 0.4$ yielded 
a converged model, but at the expense of a somewhat different model 
prescription. We show the results of this model in Figure \ref{damp25}.

The models illustrated in Figures \ref{damp15}, \ref{damp20}, and \ref{damp25}
show that in the unlikely circumstance that a star is near the base of 
the RGB at the epoch of observation, the damping masses in the envelope
remain substantially larger than the envelope mass throughout the bulk
of the envelope. These waves may make it to near the surface. These figures 
also show that the low-frequency waves of shell helium burning may escape 
diffusive damping in the envelope. For the 15 \msun\ model in the stage of 
shell carbon burning, $M_{damp}$ for waves from the helium shell remains 
well above 1 \msun\ to the outer reaches of the envelope. The damping mass 
plummets there, a point we return to below. The 20 \msun\ model 
shows a peak in $M_{damp}$ for the helium shell in the outer portions of 
the envelope, but $M_{damp}$ is substantially less than 1 \msun\ over 
much of the envelope. The 25 \msun\ model shows that $M_{damp}$ for the
helium shell exceeds the mass of the envelope through much of the envelope, 
but with a minimum that dips to close to the envelope mass. Acoustic waves 
from a star that reflected this model's properties might make it to the surface.
This model had a decreased rate of mass loss compared to the lower-mass models. 
Whether adjustment of mass loss or other model parameters would yield more 
dissipation in the 25 \msun\ or less in the 20 \msun\ model requires further 
investigation. For all three models, the high-frequency waves from carbon 
shell burning are strongly damped in the extended, dilute envelope 
corresponding to the tip of the RGB and are unlikely to make it to the 
surface despite the greater power generated compared to the helium 
shell. 

%We note again that depending on the definition of wave frequency in 
%Equation \ref{time}, the frequency could be lower and hence the damping 
%mass higher for a given values of $v_{conv}$ and $H_p$ by a factor of 
%$(2 \pi)^2 \approx 40$ compared to \citet{Fuller17}. 
It is also important 
to keep in mind that while we have estimated characteristic frequencies, 
the inner convective regions will each produce a broad spectrum of acoustic 
frequencies. Some of these will be more prone to dissipation, some less so.  

\begin{figure}%[htp]
\center
\includegraphics[width=3 in, angle=0]{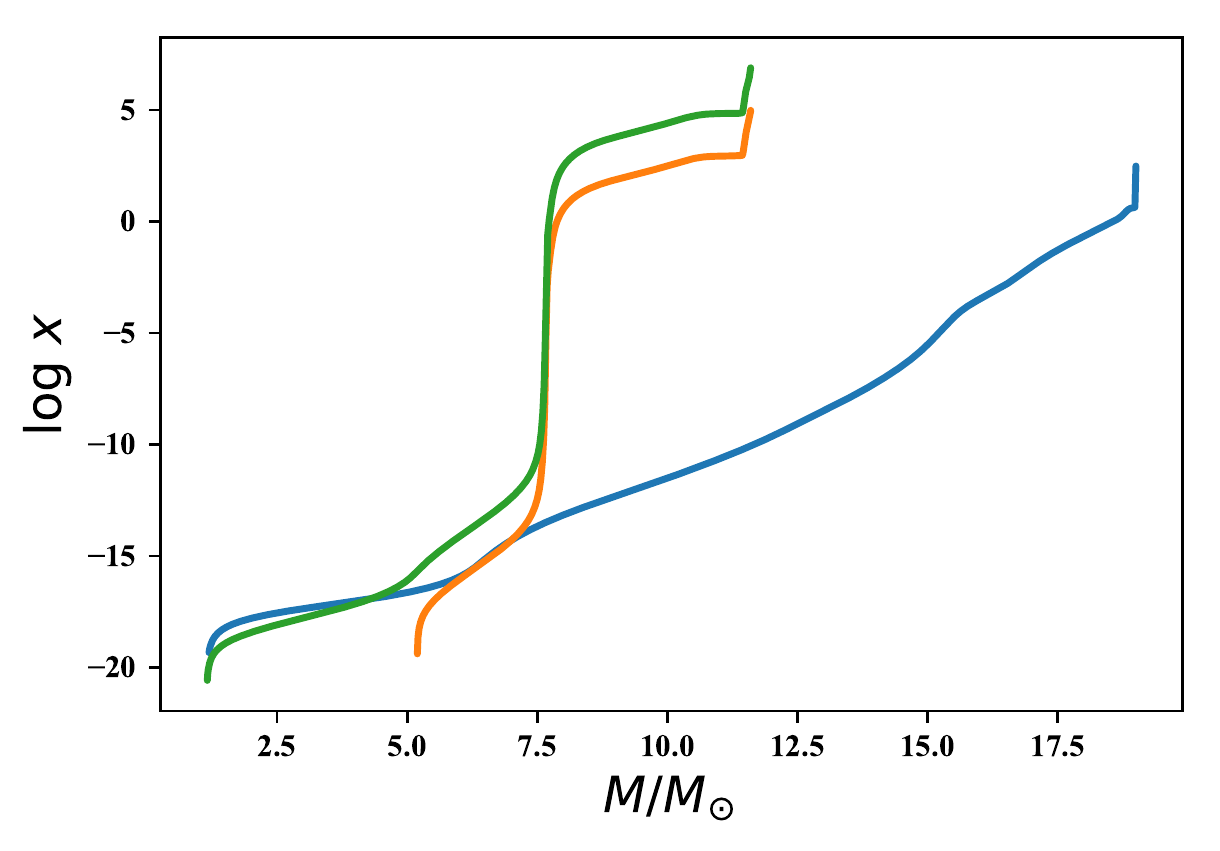}
\caption
{The quantity $x = \int \frac{dm}{M_{damp}} \equiv 2d_a$ is presented as 
a function of mass for the non-rotating 20~\msun\ model for the convective 
helium core at the base of the RGB (blue line), the helium shell in carbon shell
burning (orange line) and the carbon shell (green line).  
\label{attenuate}}
\end{figure}

Another perspective on thermal dissipation can be obtained by taking
\begin{equation}
\frac{dL}{m} = - \frac{L}{M_{damp}}
\label{diss}
\end{equation}
\citep{Fuller17}, so we can write
\begin{equation}
ln\frac{L}{L_0} = -\int \frac{dm}{M_{damp}} 
\label{perturb}
\end{equation}
where $L_0$ is the generated acoustic power and $L$ is evaluated
at a given mass coordinate. The energy in the waves is thus dissipated
by a factor $exp(-2d_a)$, and the amplitude of the waves by a factor
$exp(-d_a)$. Figure \ref{attenuate} gives the quantity, $2d_a$, as a 
function of mass for the non-rotating 20 \msun\ model, again in core 
helium burning at the base of the RGB and in the carbon shell-burning 
phase. This attenuation is less than unity for most of the core helium 
burning model until the very outer portions of the envelope where it might 
somewhat exceed unity in the amplitude, consistent with Figure \ref{damp20} 
(neglecting the very outer spike for now). Table \ref{tab:delta} also gives the 
log of the amplitude of the dissipation factor, $d_a$, for three relevant cases evaluated 
in the outer envelope but prior to the very outer spike in $M_{damp}$ for
the non-rotating 20 \msun\ model. While the small amplitudes of the acoustic 
waves from the early core helium burning may be dissipated by a factor of 
order 10, the waves from both the helium and carbon shells in the later phases 
are severely dissipated in the 20 \msun\ model, even before encountering the 
spike in $M_{damp}$. We note from Figures \ref{damp15}  and \ref{damp25} 
that the dissipation of acoustic waves from the helium-burning shell is especially 
severe for the 20 \msun\ model and that the attenuation for those waves may be 
less for other masses. 

\subsubsection{Effects of Outer Envelope}
\label{outer}

\begin{figure}%[htp]
\center
\includegraphics[width=3 in, angle=0]{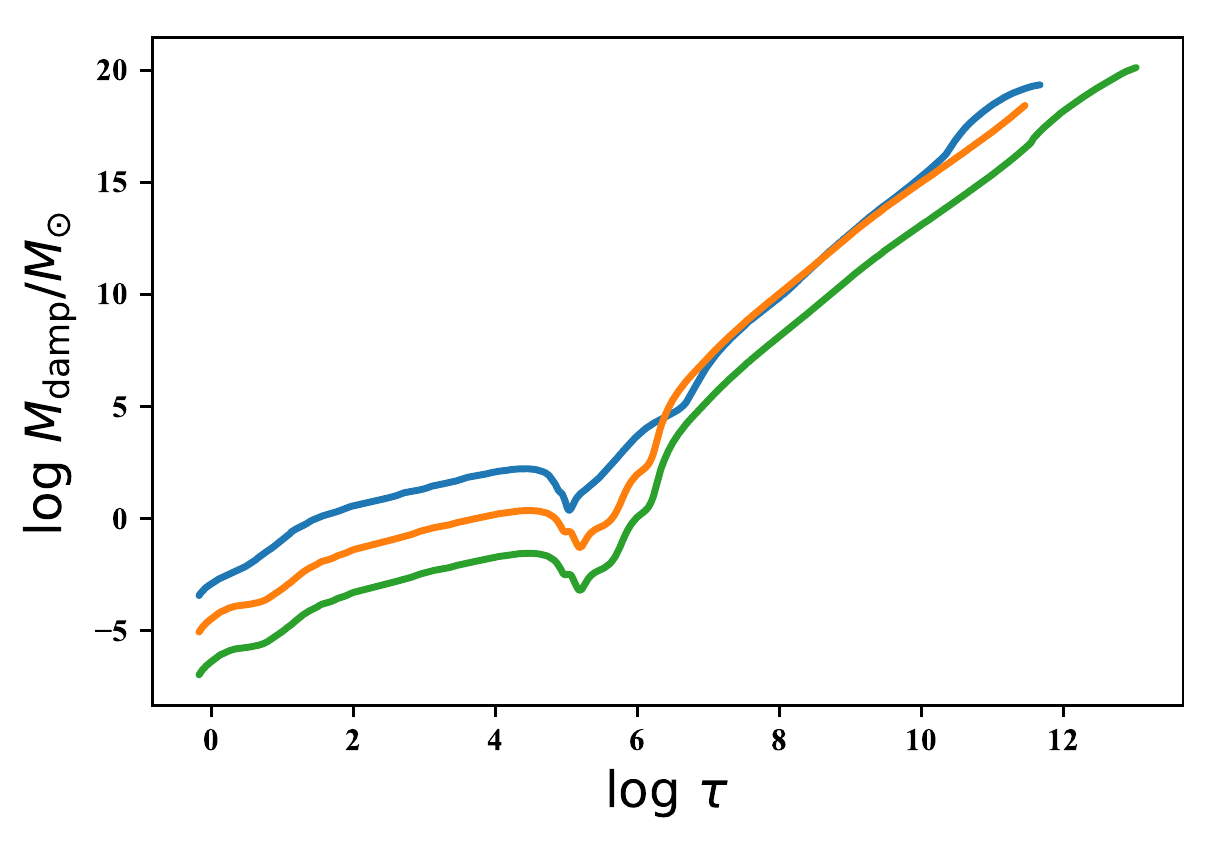}
\caption
{The damping mass for thermal diffusion at a given radius is given as  function of optical
depth from that position to the surface for the non-rotating model of 20~\msun\ for the 
convective helium core at the base of the RGB (blue line), the helium shell in carbon shell
burning (orange line) and the carbon shell (green line).
\label{tau}}
\end{figure}

To this point, we have ignored the behavior of the wave amplitude, the 
damping mass, and the attenuation factor in the very outermost layers, where 
Figures \ref{damp15}, \ref{damp20}, and \ref{damp25}, all show a peak and 
then a steep drop and Figures \ref{amplitude} and \ref{attenuate} show a spike 
for all three cases illustrated. Concerned that this was some numerical issue, 
we examined this outer structure more carefully. To this end, we present in 
Figure \ref{tau} a plot of $M_{damp}$ as a function of optical depth from the
surface for the same three cases represented in Figure \ref{attenuate}. This 
shows that the peak in $M_{damp}$ in Figures \ref{damp15}, \ref{damp20}, 
and \ref{damp25} occurs at large optical depth, $\tau \sim 10^{4.5}$, not in 
the outer atmosphere. For reference, the small outer ledge in $M_{damp}$ 
that can be discerned in Figures \ref{damp15} through \ref{damp25} occurs
 at $\tau \sim 20$. 

The outer peak in $M_{damp}$ typically occurs at a mass depth of about
1 \msun\ from the surface. The damping mass rapidly plummets outward
from the peak. The outer steep declines in $M_{damp}$ in Figures \ref{damp15}, 
\ref{damp20}, and \ref{damp25} and the outer spikes in Figures \ref{amplitude} 
and \ref{attenuate} are real and ubiquitous. These features are associated with 
the strong negative gradient in $\rho$ (see equation \ref{damp}) in the outer, 
but still optically-thick layers. This implies that even if acoustic waves reached 
the location of the outer peak in $M_{damp}$, they are likely to thermally 
dissipate before reaching the surface. 

\section{Discussion}
\label{discuss}

There is potential to determine the mass, the rotation, the internal
structure, and the evolutionary state of Betelgeuse and other red supergiants, 
if any faint, fast, acoustic perturbations could be detected. By the time 
the models have reached the tip of the giant branch, the outer, extended 
convective envelope is well established, but the inner structure is 
quite variable, with convective regions developing and vanishing and 
with different configurations at different times for different ZAMS masses. 
That Betelgeuse is in some phase of core helium burning is most probable, 
but later stages are not precluded. The inner convective regions should 
generate gravity waves that convert to acoustic waves with characteristic 
periods of 20 days in core helium burning, 10 days in helium shell burning, 
and 0.1 to 1 day in shell carbon burning.

Whereas \citet{SQ14} and \citet{Fuller17} focused on very late stages where
acoustic flux could be strong enough to drive envelope expansion or 
mass loss, here we sought conditions where acoustic waves might
give some diagnostic of the inner properties of red supergiants like 
Betelgeuse. \citet{SQ14} and \citet{Fuller17} caution that dissipation may 
prevent any acoustic waves from propagating to the surface. Although our 
models in this survey are a sparse sample of the evolutionary states, we 
largely confirm these pessimistic results. We find that acoustic waves may 
propagate to near the surface, avoiding both shock and diffusive dissipation 
relatively early in core helium burning with a luminosity amplitude of about a 
millimagnitude. By the time the models are in shell carbon burning, the signals 
leaving the helium and carbon shells have more power, but shock and 
especially thermal dissipation and attenuation cause damping in the
outer convective envelope. This damping may be less for acoustic waves
generated by the helium-burning shells in the the 15 \msun\ and 25 \msun\ 
models than for the 20 \msun\ model, but it is severe in all the models for
the waves from the carbon-burning shell.

In all the models we investigated, the rapid decline in density in the outermost, 
but still optically thick, layers, seemed to provide a last blockade against 
observable perturbations to the surface luminosity, dissipating any impinging 
acoustic wave by some combination of thermal dissipation and shock formation. 
Other factors are neutrino losses in the core or wave reflection, both of which 
we have ignored here.

We have presented rudimentary theory here that is subject to
manifest uncertainties. Betelgeuse and most other massive RSG
are most likely to be in mid to late core helium burning -- the
longest-lived phase after the main sequence -- for which
we predict little observable signal. On the other hand if 
Betelgeuse were in core silicon burning with only days to 
live, there would be no time for waves to reach the surface
and hence little to see \citep{Fuller17}. Absence of evidence 
does not mean things might not get exciting very quickly. 

It would be interesting to do a more careful study of structure while 
near the tip of the RSB. Of interest would be to establish the phase 
of core helium burning when evidence for that core convection first 
becomes observable in principle, with the convective frequency greater 
than the envelope cutoff frequency and to probe the evolving structure 
at and beyond core helium burning. The nature of the inner convective 
regions, their characteristic frequencies and power, and how they vary 
in time, require a dedicated study. We have not explored basic model
sensitivity to issues like mixing length, overshoot, etc. We have done
a basic exploration of the effects of rotation and associated mixing. Taking 
the possible effects of magnetic dynamos into account would add more 
complexity, but is within the capabilities of $MESA$. Even if acoustic wave 
perturbations cannot be observed directly, there may be some observable 
change in the structure of the outer envelope, its radius or its mass loss.

As mentioned in the Introduction, another potentially important constraint
on Betelguese specifically is the observed 420 day pulsation mode. The 
pulsation period for convective envelopes scales as $P \propto R^2/M$ 
(Gough et al. 1965), so the period provides an important independent 
constraint on $R$ and $M$, depending on the proportionality constant. 
From the models of Heger et al. (1997), the envelope pulsation of 
Betelgeuse is most likely to correspond to a first overtone. For its 
mass and luminosity, the fundamental period for Betelgeuse would be 
several times larger than the observed period. In these models, the 
pulsation grows to a non-linear limit and then has the potential to 
expel a super wind \citep{yoon10}. Betelgeuse is not doing that. 
It remains of great interest to try to establish the conditions, that 
may depend on rotation, magnetic fields, and whether a companion has 
been ingested, that will yield a stable non-linear pulsation 
with the observed period.

\citet{SQ14} and \citet{Fuller17} have argued that dissipation of acoustic
waves in the envelope may lead to expansion of the envelope and mass loss
of various intensities. There is evidence that Betelgeuse has a region of
extended material out to perhaps 10 stellar radii beyond the photosphere
\citep{Kervella11} and that upward and downward flows extend into that
material \citep{Ohnaka11}. \citet{Dessart17} argue that this material
may routinely affect the breakout properties of a supernova shock. It is
possible that this outer, dynamical shell is the result of dissipation
of acoustic energy from the inner convective regions. If this is so,
then it may be that there is no asteroseismological signal to be 
seen from Betelgeuse because the inevitable waves have been dissipated
to expand the envelope. 

On the other hand, it may be that the propagation
of acoustic waves is more complex than has yet been explored. For instance,
there may be percolation properties whereby outgoing acoustic waves
selectively permeate cooler, denser, down-flowing plumes on their
inward journey. Even if acoustic waves percolate, their effect on
the surface might average out over the surface. Perhaps individual 
surface plumes would need to be studied for their asteroseismological 
properties. 

We have treated the acoustic flux as if it had a single frequency
from a single convective region, but every convective region will emit a
spectrum of waves; perhaps some wavelengths are more penetrating
that others. Another possibility might be to examine the surface properties
in different wavelengths where the opacity is less (E. Levesque,
private communication, 2018). 

We have also seen suggestions that some masses may be more amenable 
to propagating acoustic waves than others. If, at some masses, one
could discern acoustic perturbations from early and later core or
shell helium burning, that stage of helium burning could be diagnosed.

It may very well be that there are no acoustic signals from the inner
convective structure of RSG that reach the surface yielding clues
to the inner structure and hence nothing to observe. Yet detection
of such perturbations could provide a rich new way to probe the
inner structure of supernova progenitors. Any detection would require
an immense amount of work to quantitatively interpret the data -- we
have barely scratched model parameter space here -- but it seems a 
shame not to try.

\section*{Acknowledgments}

We are grateful to Pawan Kumar, Mike Montgomery, and Jim Fuller for discussions 
of stellar wave propagation, to David Branch, Emily Levesque, and Brad Schaefer 
for perspectives on red supergiant evolution, and to Brian Mulligan for LaTeX advice. 
We are especially thankful for the ample support of Bill Paxton and the MESA team. 
This research was supported in part by NSF Grant AST-11-9801 and in part by the 
Samuel T. and Fern Yanagisawa Regents Professorship in Astronomy.

%%%%REFERENCES%%%%%%%

\end{document}

%% file: definitions.tex
\def\la{\mathrel{\hbox{\rlap{\hbox{\lower4pt\hbox{$\sim$}}}\hbox{$<$}}}}
\def\ga{\mathrel{\hbox{\rlap{\hbox{\lower4pt\hbox{$\sim$}}}\hbox{$>$}}}}

\def\kms{km~s$^{-1}$}
\def\cms{cm~s$^{-1}$}

\def\dm15{{$\Delta$}$m_{15}$}

\def\v10{$V_{10}$(Si~II)}

\def\W575{$W(5750)$}
\def\W610{$W(6100)$}
\def\6100{the 6100~\AA\ absorption}

\def\m{M$_\odot$}
\def\msun{M$_\odot$}

\def\ergs{erg~s$^{-1}$}
\def\gcm3{g~cm$^{-3}$}
\def\cm2g{cm$^{2}$~g$^{-1}$}

\def\CaII7291{[Ca~{\sc II}] $\lambda\lambda$7291,7323}

\def\OI6300{[O~{\sc I}] $\lambda\lambda$6300,6364}

\def\apj{ApJ}
\def\apjl{ApJL}
\def\apjs{ApJS}
\def\aj{AJ}

\def\mnras{MNRAS}
\def\aap{A\&A}